\documentclass[%
 aip,
 jmp,%
 amsmath,amssymb,
 reprint,%
]{revtex4-1}

\usepackage{graphicx}
\graphicspath{{Figures/}}
\usepackage{dcolumn}
\usepackage{bm}
\usepackage{subfigure}
\usepackage{comment}
\usepackage{multirow}
\usepackage{listings}
\usepackage[normalem]{ulem}
\usepackage{algpseudocode}
\usepackage{amsmath,amssymb,amsfonts}
\usepackage{graphics}
\usepackage{color}
\usepackage{mathtools}

\usepackage[ruled,vlined]{algorithm2e}

\definecolor{rev1}{rgb}{0,0,0}

\definecolor{backcolour}{rgb}{0.95,0.95,0.92}
\lstdefinestyle{mystyle}{
    frame=tb,
    basicstyle=\small,
    breakatwhitespace=false,
    breaklines=true,
    captionpos=b,
    keepspaces=true,
    numbers=left,
    xleftmargin=2em,
    numbersep=5pt,
    showspaces=false,
    showstringspaces=false,
    showtabs=false,
    tabsize=5,
}
\begin{document}


\title{A deep learning enabler for non-intrusive reduced order modeling of fluid flows} 

\author{S. Pawar}
\author{S. M. Rahman}
\author{H. Vaddireddy}
\author{O. San}%
\email{osan@okstate.edu}
\affiliation{ 
School of Mechanical \& Aerospace Engineering, Oklahoma State University, Stillwater, Oklahoma - 74078, USA.
}%


\author{A. Rasheed}
\affiliation{%
Department of Engineering Cybernetics, Norwegian University of Science and Technology, N-7465, Trondheim, Norway.
}%

\author{P. Vedula}
\affiliation{%
School of Aerospace \& Mechanical Engineering, The University of Oklahoma, Norman, Oklahoma - 73019, USA.
}%

\date{\today}

\begin{abstract}
In this paper, we introduce a modular deep neural network (DNN) framework for data-driven reduced order modeling of dynamical systems relevant to fluid flows. We propose various deep neural network architectures which numerically predict evolution of dynamical systems by learning from either using discrete state or slope information of the system. Our approach has been demonstrated using both residual formula and backward difference scheme formulas. However, it can be easily generalized into many different numerical schemes as well. We give a demonstration of our framework for three examples: (i) Kraichnan-Orszag system, an illustrative coupled nonlinear ordinary differential equations, (ii) Lorenz system exhibiting chaotic behavior, and (iii) a non-intrusive model order reduction framework for the two-dimensional Boussinesq equations with a differentially heated cavity flow setup at various Rayleigh numbers. Using only snapshots of state variables at discrete time instances, our data-driven approach can be considered truly non-intrusive, since any prior information about the underlying governing equations is not required for generating the reduced order model. Our \textit{a posteriori} analysis shows that the proposed data-driven approach is remarkably accurate, and can be used as a robust predictive tool for non-intrusive model order reduction of complex fluid flows.
\end{abstract}

\keywords{Dynamical systems, Deep learning, Neural networks, Reduced order modeling, Proper orthogonal decomposition} 
\maketitle


\section{Introduction} 
\label{sec:intro}

Many realistic transient flows typically involve very wide ranges of spatial and temporal scales, which place an enormous computational burden on direct numerical simulations (DNS) of such flows based on governing equations. Advancement in high-performance computing systems along with the development of consistent, stable, convergent numerical schemes, and efficient parallel algorithms has enabled us to analyze and study complex real world processes. For instance, it is now possible to collect very high-resolution DNS data relevant to selected turbulent flows which cannot be gathered experimentally \cite{ishihara2009study}. However, the computational cost of performing DNS scales roughly as $\text{Re}^3$, where Re is the Reynolds number of the flow \cite{karniadakis1993nodes}. Hence, with the present state of the art computing architectures \cite{reed2015exascale}, such high-resolution simulations of multiphysics flows might require weeks of computations even for simple geometries. The situation worsens when a series of numerical simulations need to be run for any parametric design optimization study. To alleviate this, coarse-graining approaches, as performed for example in large eddy simulations (LES), are commonly used to reduce this computational burden \cite{mason1994large,piomelli1999large,meneveau2000scale,moin2002advances}.  

However, the computational cost of full-order simulations (i.e., DNS or even LES) can still be considered extremely prohibitive due to a large number of degrees of freedom needed to resolve all of the flow features, especially in settings where the traditional methods require repeated model evaluations over a large range of parameters. Therefore, many successful model order reduction approaches have been introduced \cite{lucia2004reduced,benner2015survey,brunton2015closed,taira2017modal,rowley2017model}. The main purpose of such approaches is to reduce this computational burden and serve as surrogate models for efficient computational analysis of fluid systems. A common objective in such reduced order modeling (ROM) approaches is to determine how well these approaches can reproduce the flow dynamics. 

\textcolor{rev1}{Intrusive finite dimensional low order models routinely arise when we apply Galerkin type projection techniques to infinite dimensional models\cite{lakshmivarahan2008relation,lakshmivarahan2008structure,wang2009relation}. On the other hand, without prior information on the governing equations, their operator forms or parameterizations to account for complex physical processes, a non-intrusive ROM approach can be reconstructed to infer such underlying physics from the data itself. Having ability to facilitate dynamic data exchange easier between different components, these non-intrusive models can arguably be more promising and impactful in numerous interdisciplinary fields. Moreover, with the advent of digital twin technologies\cite{8477101}, the collection of data from sensors has become possible at different stages of product's lifecycle, and model order reduction might be considered as a key enabler for this digital twin vision in many emerging cyber-physical systems\cite{hartmann2018model}.}

Reduced order models offer promises in many fields such as system identification \cite{juang1985eigensystem,harish2016reduced,fung2019multiscale}, control \cite{ito1998reduced,kunisch1999control,hovland2008explicit,bergmann2008optimal,noack2011reduced,karcher2018certified}, optimization \cite{lassila2010parametric,benner2014model,heinkenschloss2018reduced}, and data assimilation \cite{cao2007reduced,daescu2008dual,cstefuanescu2015pod} applications. 
In these model reduction approaches, we aim at obtaining simplified (but dense) models from high-fidelity numerical simulation data or data collected from the experiment \cite{rowley2017model, jolliffe2016principal}. To fulfill their objectives in multiple forward simulations of the problem with different model parameters \cite{allen2004application,frangos2010surrogate,degroote2010interpolation,weickum2006multi,shenefelt2002solution,lieberman2010parameter}, these models should be sufficiently accurate and computationally much faster than the high-fidelity numerical simulation. Therefore, there has been progress made in recent years to develop such ROM approaches specifically for nonlinear systems \cite{everson1995karhunen, chaturantabut2010nonlinear, noack2003hierarchy, carlberg2011efficient, alla2017nonlinear, loiseau2018sparse}.

The basic philosophy of projection-based ROM approaches is to reduce the high degrees of freedom of a governing equation through an expansion in a transformed space, traditionally with orthogonal basis. Among the large variety of projection-based ROM strategies, the proper orthogonal decomposition (POD) has emerged as a popular technique for the study of dynamical systems \cite{berkooz1993proper,antoulas2000survey, benner2015survey, rowley2017model}, which targets the most dominant characteristics of the flow considering the largest energy containing modes. The POD technique was first introduced in fluid community in the context of extracting coherent structure from turbulent flow field \cite{lumley1967structure}. Several methods have also been proposed in literature aimed at improving the POD modes \cite{bui2007goal,hay2009local,kunisch2010optimal,carlberg2011low}. There are also different variants of POD that have been introduced, like spatio-temporal biorthogonal decomposition \cite{aubry1991spatiotemporal}, spectral POD (SPOD)\cite{sieber2016spectral}, frequency based POD that is also called as SPOD\cite{towne2018spectral}, multiscale POD (MPOD)\cite{mendez2019multi} which splits the correlation matrix into the contribution of different scales.

The evolution equations for the lower order system are then obtained using the Galerkin projection method. For many flows, the POD-Galerkin method provides an efficient and accurate way to generate ROM methodologies \cite{aubry1988dynamics, rowley2004model, barone2009stable, akhtar2009stability, kalb2007intrinsic,bergmann2009enablers, ravindran2000reduced, amsallem2009method, san2015principal}. Furthermore, several successful closure models have been suggested in order to model the effects of discarded modes \cite{borggaard2011artificial,wang2012proper,san2014proper,osth2014need}. \textcolor{rev1}{The POD-Galerkin intrusive approach can also be stabilized with a nonlinear eddy viscosity model\cite{cordier2013identification} or with proper selection of linear quadratic coefficients\cite{schlegel2015long}.} However, the projection-based model reduction approaches have limitations especially for complex systems such as general circulation models, since there is a lack of access to the full-order model operators or the complexity of the forward simulation codes that render the need for obtaining the full-order operators \cite{lassila2014model,peherstorfer2016data,siddiqui2019finite}.

\textcolor{rev1}{One of the challenges in Galerkin projection is the deformation of POD modes. As recently discussed by \citeauthor{reiss2018shifted}\cite{reiss2018shifted}, transport-dominated phenomena are usually a challenge for modal methods, since their dynamics cannot be captured accurately by a few dominant spatial modes. If we include more number of modes to better recover the embedded structures in the underlying system, the computational expense increases and the ROM might not be efficient from practical point of view. Furthermore, the construction of a least-order state space is crucial especially in control since every degree of freedom can amplify noise\cite{ehlert2019locally}. A data-driven manifold learning model has been proposed as a general dynamic ROM modeling framework\cite{loiseau2018sparse}. \citeauthor{ehlert2019locally}\cite{ehlert2019locally} have also presented a manifold representation of the transient oscillatory cylinder wake using a locally linear embedding approach as encoder. They found that this representation outperforms a 50-dimensional POD expansion from the same data. Another key dynamic problem is that hyperbolic convection problems are treated with an elliptic Galerkin method\cite{noack2016snapshots}. Even though the flow has certain specific direction, the Galerkin projection assumes that the modes are globally coupled. This mismatch between Navier-Stokes equations and Galerkin dynamics might not be curable. Also, the frequency range of high-dimensional Navier-Stokes solutions is not resolved in the low-dimensional deterministic system\cite{aubry1988dynamics,bourgeois2013generalized}. \citeauthor{rempfer2000low}\cite{rempfer2000low} has demonstrated some implications of projecting the Navier-Stokes equations onto low-dimensional bases and showed how the restriction to a low-dimensional basis as well as improper treatment of boundary conditions might affect the validity of ROM.} Due to all these limitations, there is a recent interest in generating fully non-intrusive approaches without the need for access to full-order model operators to establish surrogate models \cite{audouze2013nonintrusive,mignolet2013review,xiao2015non,xiao2015non,peherstorfer2016data,hesthaven2018non,hampton2018practical,chen2018greedy,xiao2019domain,wang2019non}. \textcolor{rev1}{Dynamic mode decomposition (DMD) models\cite{rowley2009spectral,schmid2010dynamic,jovanovic2014sparsity,hemati2014dynamic,proctor2015discovering,noack2016recursive} provide this non-intrusive representation directly from data by their nature, and several approaches have been readily available for optimal mode selection\cite{bistrian2015improved,alekseev2016linear,bistrian2017randomized,pascarella2019adaptive}.}

There is a broad range of opportunities for the application of machine learning algorithms to develop non-intrusive reduced order models. A good discussion on the application of data-driven methods for dynamical systems can be found in a book by \citeauthor{brunton2019data}\cite{brunton2019data}. We also refer to a recent review article \cite{brunton2019machine} for a comprehensive overview of the machine learning literature in fluid mechanics. A number of studies have been done to apply data-driven techniques to predict the high-dimensional complex dynamical systems \cite{san2018extreme, xie2018data, raissi2018multistep, chen2018neural, rudy2019data, kani2017dr,yeo2019deep, qin2019data, rahman2018hybrid}. \textcolor{rev1}{\citeauthor{san2018neural}\cite{san2018neural} proposed a methodology to account for the effects of truncated POD modes using a single layer feed-forward neural network.} A multistep neural network was proposed to identify nonlinear dynamical system from the data by combining classical numerical analysis techniques with the powerful nonlinear approximation capability of neural networks \cite{raissi2018multistep}. \citeauthor{xie2018data}\cite{xie2018data} used the multistep neural network to approximate the full order model projected on low-dimensional space with a supervised learning task. A deep residual recurrent neural network was introduced as an efficient model reduction technique for nonlinear dynamical systems \cite{kani2017dr}. \citeauthor{vlachas2018data}\cite{vlachas2018data} developed the data-driven forecasting method for high-dimensional, chaotic systems using the hybrid approach which combines mean stochastic model and recurrent long short-term memory (LSTM) neural network. The LSTM recurrent neural network was used to model the temporal dynamics of turbulence in a ROM framework \cite{mohan2018deep}. \citeauthor{pathak2018hybrid}\cite{pathak2018hybrid} proposed a hybrid forecasting model combining the knowledge of the governing equation of the dynamical system and machine learning technique to predict the long term behavior of chaotic systems. This hybrid approach was found to be better than either its pure data-driven component or its model-based component.    

In our proposed ROM framework, we will bypass the Galerkin projection step of the projection based ROM with our proposed neural network architectures to build a fully non-intrusive approach. This non-intrusive ROM (NIROM) framework can be viewed as a decomposition of the problem into basis representation and forecasting subproblems. We illustrate our NIROM approach using the deep feed-forward neural network architectures. However, it can be easily applied to other types of neural networks (as demonstrated for data-driven forecasting of dynamical systems \cite{vlachas2018data, lu2018beyond}) or more traditional time series forecasting tools \cite{shumway2017time}. The neural networks are capable of approximating the nonlinear functions and have been successfully used in turbulence modeling\cite{ling2016reynolds, milano2002neural, gamahara2017searching}, solving differential equation\cite{raissi2019physics, michoski2019solving}. We learn the dynamics of the reduced order model directly from the output of the full order model projected on the low-dimension space using a supervised learning task. The main advantage of this non-intrusive approach is that it does not require the information about the equations governing the full order model. 
\textcolor{rev1}{Although the proposed approach helps to generate a NIROM framework solely from the snapshot data reconstructed onto a POD-spanned space, it may still suffer from fundamental challenges of traditional POD-Galerkin models (e.g., we refer to \citeauthor{zerfas2019continuous}\cite{zerfas2019continuous} for a recent discussion about ways to mitigate their lack of accuracy).}

The paper is organized as follows: Section \ref{sec:problem_setup} introduces deep neural network architecture and implementation of different DNN frameworks for dynamical systems. Section \ref{sec:dynsystem} gives numerical results using our DNN frameworks for two dynamical systems: Kraichnan-Orszag system and Lorenz system. We present the generalized non-intrusive ROM framework in Section \ref{sec:nirom}. In Section \ref{subsec:boussinesq}, we present the Boussinesq equation problem and its POD analysis. The non-intrusive ROM framework for the Boussinesq equation is described in Section \ref{subsec:ni_rom_boussinesq}. The numerical results in Section \ref{sec:rom_results} demonstrate the effectiveness of our non-intrusive approach for reduced order modeling of differentially heated cavity problem at two Rayleigh numbers. We give some concluding remarks and suggestions for future work in Section \ref{sec:conclusion}.      

\section{Learning Framework} 
\label{sec:problem_setup}
\begin{figure*}[htbp]
\centering
\mbox{
\subfigure{\includegraphics[width=1\textwidth]{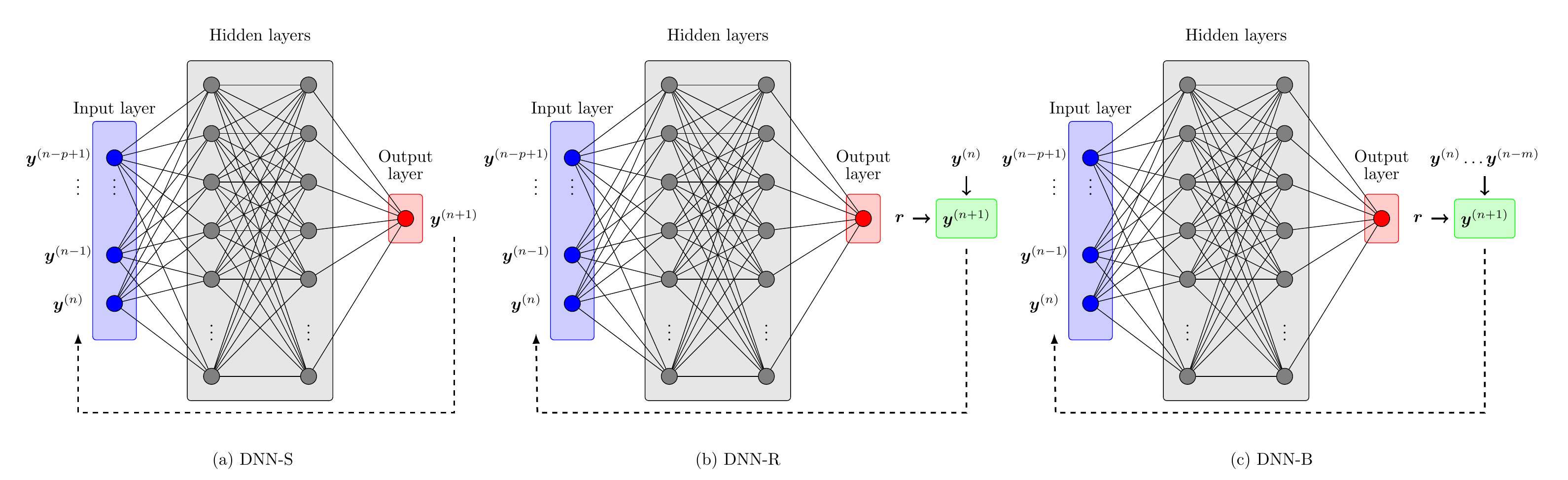}}}
\caption{Iterative prediction using the trained DNN model for different frameworks proposed in this study. The input to the neural network consists of state history for $p$ time steps. For the DNN-R framework, $\boldsymbol{r}$ is the residual predicted by the neural network. For the DNN-B framework, $\boldsymbol{r}$ is the discrete numerical slope computed using any of the numerical method used for training the neural network. The number of inputs to compute numerical slope in DNN-B framework depends upon the numerical scheme applied for calculating the slope (for example, $m=1$ for the second-order backward difference scheme used in this study).}
\label{fig:dnn_framework}
\end{figure*}
The deep neural network is an artificial neural network composed of several layers made up of the predefined number of nodes. These nodes are also called neurons. A node combines the input from the data with a set of coefficients called weights. These weights either amplify or dampen the input and thereby assign the significance to the input with respect to the output that the DNN is trying to learn. In addition to the weights, these nodes have a bias for each input to the node. The input-weight product and the bias are summed and the sum is passed through a node's activation function. The above process can be described using the matrix operation as given by \cite{hagan1996neural}
\begin{equation}
    S^l = \mathbf{W}^lX^{l-1},
\end{equation}
where $X^{l-1}$ is the output of the $(l-1)$th layer, $\mathbf{W}^l$ is the matrix of weights for the $l$th layer. The output of the $l$th layer is given by
\begin{equation}
    X^l = \zeta(S^l+B^l),
\end{equation}
where $B^l$ is the vector of biasing parameters for the $l$th layer and $\zeta$ is the activation function. If there are $L$ layers between the input and the output, then the mapping of the input to the output can be derived as follow
\begin{equation}
    Y=\zeta_L(\mathbf{W}^L,B^L,\dots,\zeta_2(\mathbf{W}^2,B^2,\zeta_1(\mathbf{W}^1,B^1,X))),
\end{equation}
where $X$ and $Y$ are the input and output of the deep neural network, respectively. 

The input layer usually takes the raw data from the training dataset and transfers it to the second layer. It does not have any biasing or activation through an activation function. The output layer usually has the linear activation function and some bias associated with the inputs. The linear activation function simply takes the summation of inputs received from the previous hidden layer and the associated bias of the output layer. In this study, we use ReLU activation function for all hidden layers and linear activation function for the output layer in all DNN frameworks. The ReLU activation function can be expressed as 
\begin{equation}
    \zeta(\chi) = \text{max}(0, \chi),
\end{equation}
where $\zeta$ is an activation function and $\chi$ is the input to the node.  

Each entry of the matrix $\mathbf{W}$ and $B$ is learned through backpropagation and some optimization algorithm. The backpropagation algorithm provides a way to compute the gradient of the objective function efficiently and the optimization algorithm gives a rapid way to learn optimal weights. The objective of the neural networks in this study is to learn the weights associated with each node in such a way that the root mean square error between the true labels $Y_0$ and output of the neural network $Y$ is minimized. The backpropagation proceeds as follows: (i) the input and output of the neural network is specified along with initial weights, (ii) the training data is run through the network to produce output $Y$ whose true value is $Y_0$, (iii) the derivative of the objective function with each of the training weight is computed using the chain rule, (iv)  the weights are updated based on the learning rate and then we go to step (ii). We continue to iterate through this procedure until convergence or the maximum number of iterations is reached. The Adams optimization algorithm \cite{shahriari2015taking} is used in this study for learning optimal weights to minimize the objective function.       

Deep neural networks are capable of approximating nonlinear dynamical systems as shown in many studies\cite{raissi2018multistep, yeo2019deep, chen2018neural, haber2017stable}. The general nonlinear dynamical system can be presented by an equation of the form
\begin{equation}
    \frac{d\boldsymbol{y}}{dt} = \boldsymbol{F}(\boldsymbol y,t),
\end{equation}
where $\boldsymbol{y}(t)$ is the state variable at time $t$ and $\boldsymbol{F}(\boldsymbol y,t)$ is the nonlinear function evaluated for each component of state variable $\boldsymbol{y}(t)$.

Figure \ref{fig:dnn_framework} shows three different DNN frameworks used in this study to predict the dynamical system. \textcolor{rev1}{Although their characteristics on the stability and well-posedness are beyond the scope of the present work, we refer to the study of \citeauthor{chang2018reversible}\cite{chang2018reversible} for several deep neural networks and their stability issues.} \textcolor{rev1}{Our motivation for different choices of DNNs comes from recent development of approximation and discovery of dynamical systems using deep learning techniques \cite{qin2019data, raissi2018multistep}. At this point, it may be inconclusive to say which DNN framework is superior as compared to others. However, our empirical evidences suggest that learning residual information or numerical slope helps in more accurate prediction of dynamical systems.} In case of the DNN-S framework, we train the neural network to learn an update formula which advances the state of the system from $\boldsymbol{y}^{(n)}$ to $\boldsymbol{y}^{(n+1)}$ directly, where $(n)$ denotes the state of the system at time $t_n$. The S in the DNN-S stands for sequential. The past history of the state of the system can be utilized to predict the system's future state by incorporating it in the input features of the neural network. If we want to include the state history for $p$ time steps, then the input of the neural network consists of $\boldsymbol{y}^{(n)}$, $\boldsymbol{y}^{(n-1)}, \dots ,\boldsymbol{y}^{(n-p+1)}$. Hence, the neural network will have $R \times p$ input features and $R$ output labels (i.e., $\mathfrak{M}: {\rm I\!R}^{R \times p} \Rightarrow {\rm I\!R}^{R}$), where $\mathfrak{M}$ refers to the DNN model, and $R$ is the number of components of the dynamical system. Once the neural network is trained and the weights are learned, the neural network is used to predict the state of the system starting with an initial condition $\boldsymbol{y}^{(0)}$ and proceeding in time iteratively. The prediction of the system in iterative fashion is shown in Figure \ref{fig:dnn_framework}a. If the neural network is trained using the state of the system for $p$ time steps, then the previous $p$ time steps should be stored in the prediction routine. The future state of the system $\boldsymbol{y}^{(n+1)}$ is predicted using the true state of the system for first $p$ time steps. After $p$ time steps, only the predicted values by the neural network are used in the input. The predicted variable of the neural network and the solution update formula during an iterative prediction are given in Table \ref{tab:tab_num_scheme}.    

For the DNN-R framework, we learn the difference between the state of the system at time step $t_n$ and next time step $t_{n+1}$. The residual between two time steps is then applied to update the current state during prediction. The learning of the residual information instead of sequential update formula helps in stabilizing the neural network\cite{haber2017stable} and also improves the accuracy of neural network prediction\cite{lu2018beyond}. The DNN-R framework can also be implemented using the history of the system's state, similar to the DNN-S framework. The related framework was employed for the model order reduction of parametric viscous Burgers equation problem\cite{SAN2019271} with one temporal leg history (i.e., $p=1$). They call it POD-ANN-RN framework and this framework was found to give better results for interpolatory and extrapolatory ROM than sequential learning. The iterative prediction of the dynamical system using DNN-R framework is shown in Figure \ref{fig:dnn_framework}b. The future state of the system is computed using the solution update formula mentioned in Table \ref{tab:tab_num_scheme}. 

We also introduce an additional framework called a DNN-B framework as shown in Figure \ref{fig:dnn_framework}c. The B in the name stands for the backward difference. In this framework, we use the second order backward difference numerical scheme to compute the slope at time step $t_n$. The numerical slope information is then used to update the state of the system during prediction. The similar approach is also implemented in other data-driven methods for dynamical systems. One such work is the Multistep neural network \cite{raissi2018multistep} used for the data-driven discovery of nonlinear dynamical systems from experimental measurements of the state of the system. In their work, the evolution of the system $\boldsymbol{F}$ is learned using the neural network by incorporating the multistep Adams-Moulton numerical scheme in computing the loss function of the neural network. In our work, we use the numerical slope as the predicted variable and use mean squared error as the loss function. One of the advantages of directly learning the numerical slope is that standard loss functions available in Keras library can be directly applied without any modification. We apply the second order backward difference scheme in DNN-B framework to compute the discrete numerical slope. However, the framework can be implemented with any family of numerical schemes such as central difference or forward difference family. The equation used to determine the numerical slope and the solution update formula during iterative prediction are provided in Table \ref{tab:tab_num_scheme}.   

The quantitative performance of each DNN framework is measured by a quantity root mean square error (RMSE). The root mean square is determined for each component $k$ between true state and the state predicted by the neural network from the initial time to final time. The root mean square of each component is added to get the total root mean square error. The root mean square error is defined as:
\begin{equation}\label{eq:rmse}
    \text{RMSE} = \sum_{k=1}^{R}\sqrt{\frac{1}{N}\sum_{i=1}^{N}\left(y_{k}^{(i)}-\tilde{y}_{k}^{(i)}\right)^2},
\end{equation}
where $R$ is the total number of components of the dynamical system, $N$ is the total number of time steps in the evolution of the dynamical system, $\boldsymbol{y}$ is the true solution and $\tilde{\boldsymbol{y}}$ is the solution predicted by the neural network.
\begin{table*}[htbp]
\centering
\caption{The output of different DNN frameworks learned through training. The trained parameter is then used to update the solution in time, starting from the initial condition.}
\begin{tabular}{>{\raggedright\arraybackslash}m{0.1\textwidth}m{0.35\textwidth}m{0.45\textwidth}}
\hline\noalign{\smallskip}
\multicolumn{1}{>{\centering\arraybackslash}m{0.1\textwidth}}{\textbf{Neural network framework}} 
    & \multicolumn{1}{>{\centering\arraybackslash}m{0.35\textwidth}}{\textbf{Predicted variable}}
    & \multicolumn{1}{>{\centering\arraybackslash}m{0.45\textwidth}}{\textbf{Solution update}} \\ \hline\noalign{\smallskip}
    
\centering DNN-S & \begin{equation*}
    \boldsymbol{r} = \boldsymbol{y}^{(n+1)}
\end{equation*} &  \begin{equation*}
    \boldsymbol{y}^{(n+1)} = \boldsymbol{r} 
\end{equation*} \\

\centering DNN-R  & \begin{equation*}
    \boldsymbol{r} =\boldsymbol{y}^{(n+1)}-\boldsymbol{y}^{(n)}
\end{equation*}  &  \begin{equation*}
    \boldsymbol{y}^{(n+1)} = \boldsymbol{y}^{(n)}+ \boldsymbol{r}
\end{equation*} \\ 

\centering DNN-B & \begin{equation*}
    \boldsymbol{r} = \frac{3\boldsymbol{y}^{(n+1)}-4 \boldsymbol{y}^{(n)} + \boldsymbol{y}^{(n-1)}}{2 \Delta t}
\end{equation*}  &  \begin{equation*}
    \boldsymbol{y}^{(n+1)} = \frac{4}{3}\boldsymbol{y}^{(n)} - \frac{1}{3}\boldsymbol{y}^{(n-1)} + \frac{2}{3} \boldsymbol{r} \Delta t 
\end{equation*} \\
\hline

\end{tabular}
\label{tab:tab_num_scheme}
\end{table*}

\section{Time Series Prediction for Dynamical Systems} 
\label{sec:dynsystem}
Before implementing our proposed DNN frameworks within the non-intrusive ROM setup, we demonstrate the capability of our DNN frameworks to model nonlinear dynamical systems using two examples. Section \ref{sec:ko} provides numerical results for three mode Kraichnan-Orszag problem and Section \ref{sec:lorenz} gives results for the chaotic Lorenz system.

\subsection{Kraichnan-Orszag System}
\label{sec:ko}
We start by considering the nonlinear dynamical system Kraichnan-Orszag \cite{wan2005adaptive} of order $R=3$ as the first test problem. The Kraichnan-Orszag system is defined as:
\begin{equation}
    \frac{dy_1}{dt} = y_1 y_3; \quad \frac{dy_2}{dt} = -y_2 y_3; \quad \frac{dy_3}{dt} = -y_1^2 + y_2^2,
\end{equation}
with an initial condition $y_1(0) = 1$, $y_2(0) = 0.1 \xi$, and $y_3(0) = 0$. The input $\xi$ lies between the interval $[-1,1]$. The modeling of this system is particularly challenging due to the discontinuity at planes $y_1(0)=0$ and $y_2(0)=0$.

The system is solved numerically using the SciPy function \texttt{odeint} for time integration between the time interval $[0,10]$ and $N$ = 1000 time steps ($\Delta t = 0.01$). The initial condition considered for the system is $\xi=0.5$ (i.e., $[y_1 \; y_2 \; y_3]^T$ = $[1.0 \; 0.05 \; 0.0]^T$). The true solution is used for training the neural network. We apply same hyperparameters for all DNN frameworks. We use three hidden layers with 128 neurons each. The maximum number of iterations is set to 600 and 10\% of the data is used for validation to avoid overfitting.  

\begin{figure*}[htbp]
\centering
\mbox{
\subfigure{\includegraphics[width=0.9\textwidth]{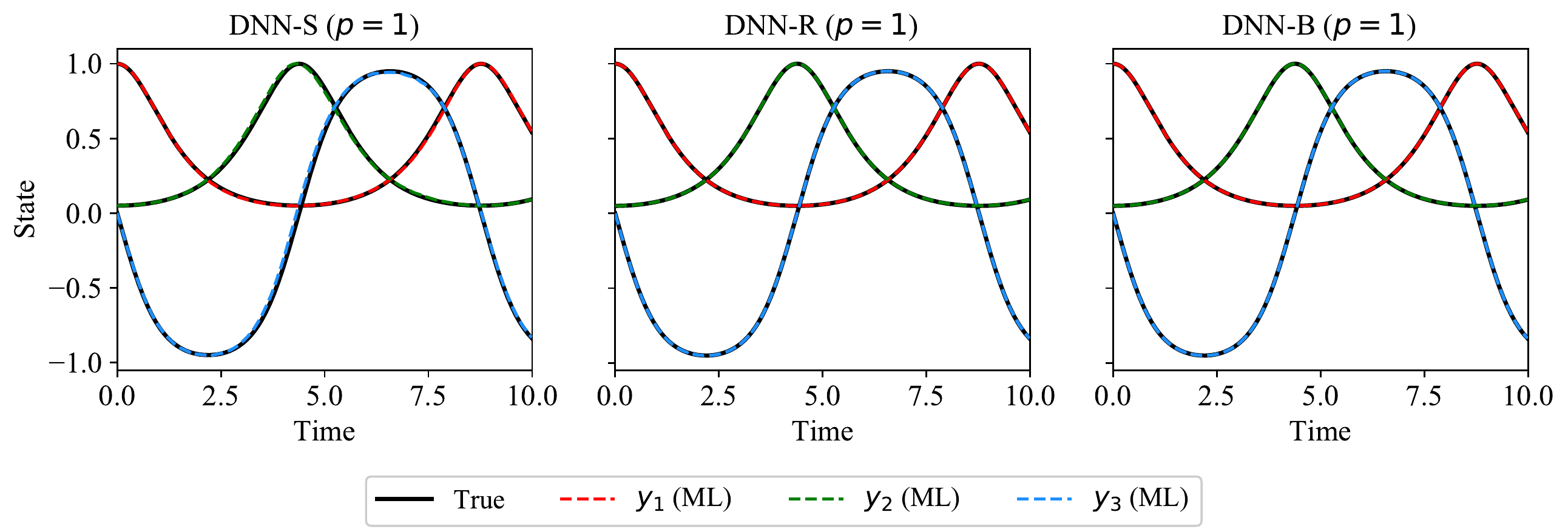}}
}
\caption{Prediction of the state of the Kraichnan-Orszag system for different DNN frameworks trained using $p=1$. The initial condition of the dynamical system is $[y_1 \; y_2 \; y_3]^T=[1.0 \; 0.05 \; 0.0]^T$. The solid lines present the true state of the system and the dashed line presents the state of the system predicted using neural network.}
\label{fig:ko_oneleg}
\end{figure*}

\begin{figure*}[htbp]
\centering
\mbox{
\subfigure{\includegraphics[width=0.9\textwidth]{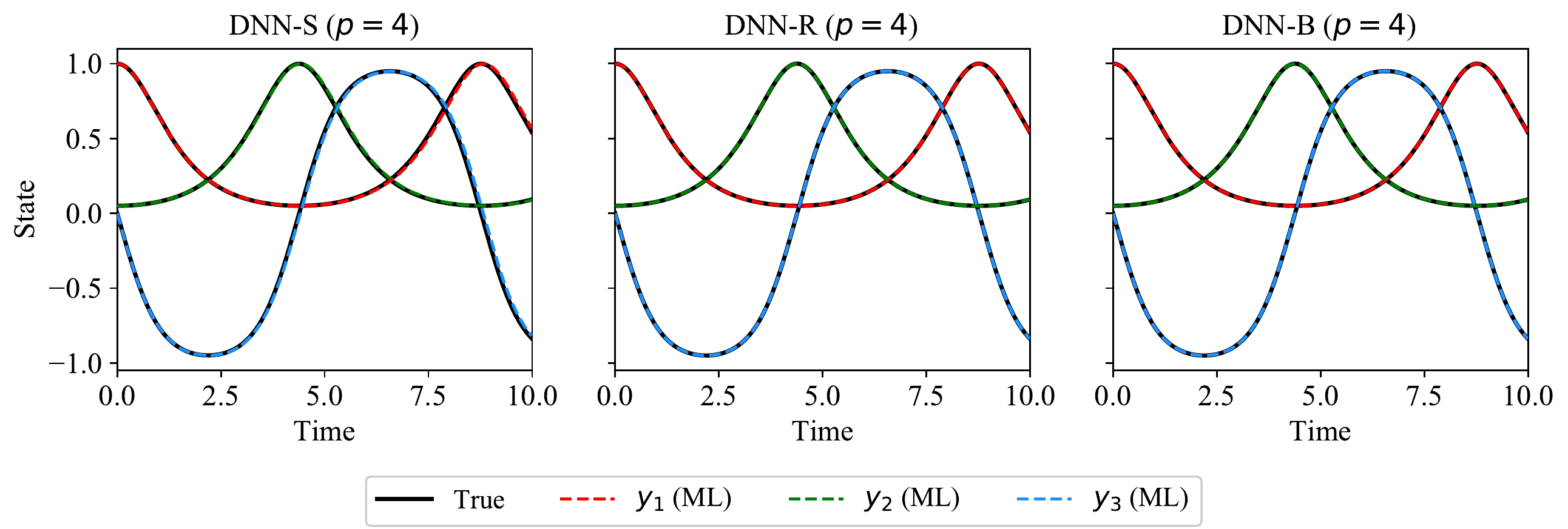}}
}
\caption{Prediction of the state of the Kraichnan-Orszag system for different DNN frameworks trained using $p=4$. The initial condition of the dynamical system is $[y_1 \; y_2 \; y_3]^T=[1.0 \; 0.05 \; 0.0]^T$. The solid lines present the true state of the system and the dashed line presents the state of the system predicted using neural network.}
\label{fig:ko_fourleg}
\end{figure*}

The true state of the system and the state predicted by different DNN frameworks are shown in Figure \ref{fig:ko_oneleg} for $p=1$. All DNN frameworks are correctly able to predict all three states of the dynamical system. We also see that DNN-R and DNN-B frameworks perform better than the DNN-S framework. For DNN-S framework, the state predicted by the neural network is very slightly shifted from the original state. Figure \ref{fig:ko_fourleg} provides the numerical results for DNN frameworks for $p=4$. The state predicted by all DNN frameworks is almost the same as the true state. Table \ref{tab:ko_rmse} compares the RMSE calculated using Equation~(\ref{eq:rmse}) for all three DNN frameworks with different number of inputs to the neural network.

 \begin{table}[htbp]
\centering
\caption{Quantitative assessment of different DNN frameworks with different number of inputs to the neural network for the Kraichnan-Orszag system using the total root mean square error given by Equation~\ref{eq:rmse}.}
\begin{tabular}{p{0.15\textwidth}p{0.14\textwidth}p{0.14\textwidth}}\\
\hline\noalign{\smallskip}
\textbf{Framework} & $\text{RMSE}~({p=1})$ & $\text{RMSE}~({p=4})$\\\hline\noalign{\smallskip}
DNN-S & $1.71 \times 10^{-2}$ & $1.90 \times 10^{-2}$ \\
DNN-R & $1.02 \times 10^{-3}$ & $2.48 \times 10^{-4}$ \\
DNN-B & $5.67 \times 10^{-4}$ & $5.97 \times 10^{-4}$ \\
\hline
\end{tabular}
\label{tab:ko_rmse}
\end{table}

\subsection{Lorenz System }
\label{sec:lorenz}
The Lorenz system\cite{lorenz1963deterministic} can be described by equations given below
\begin{equation}
\begin{split}
    \frac{dy_1}{dt} &= \alpha(y_2-y_1), \\
    \frac{dy_2}{dt} &= y_1(\rho - y_3) - y_2, \\
    \frac{dy_3}{dt} &= y_1 y_2 -\beta y_3.
\end{split}    
\end{equation}
For the Lorenz system, we use $\alpha=10$, $\rho=28$, and $\beta=8/3$. The modeling of the dynamics of the Lorenz system is a challenging problem due to its highly nonlinear and chaotic behavior. This system arises in many simplified models for physical processes \cite{berg1986order, poland1993cooperative}. 

The training data for the neural network is generated using the true solution to the Lorenz system. We use the initial condition $[y_1 \; y_2 \; y_3]^T = [-8 \; 7 \; 27]^T$ and generate the true solution numerically using the SciPy function \texttt{odeint}. The true solution is generated between the time interval $[0,25]$ with the time step $\Delta t = 0.01$ ($N = 2500$ time steps). We apply the same hyperparameters for all DNN frameworks. We use five hidden layers with 128 neurons each. The maximum number of iterations is set to 1000 and 10\% of the training data is used for validation to avoid overfitting.   

\begin{figure*}[htbp]
\centering
\mbox{
\subfigure{\includegraphics[width=0.9\textwidth]{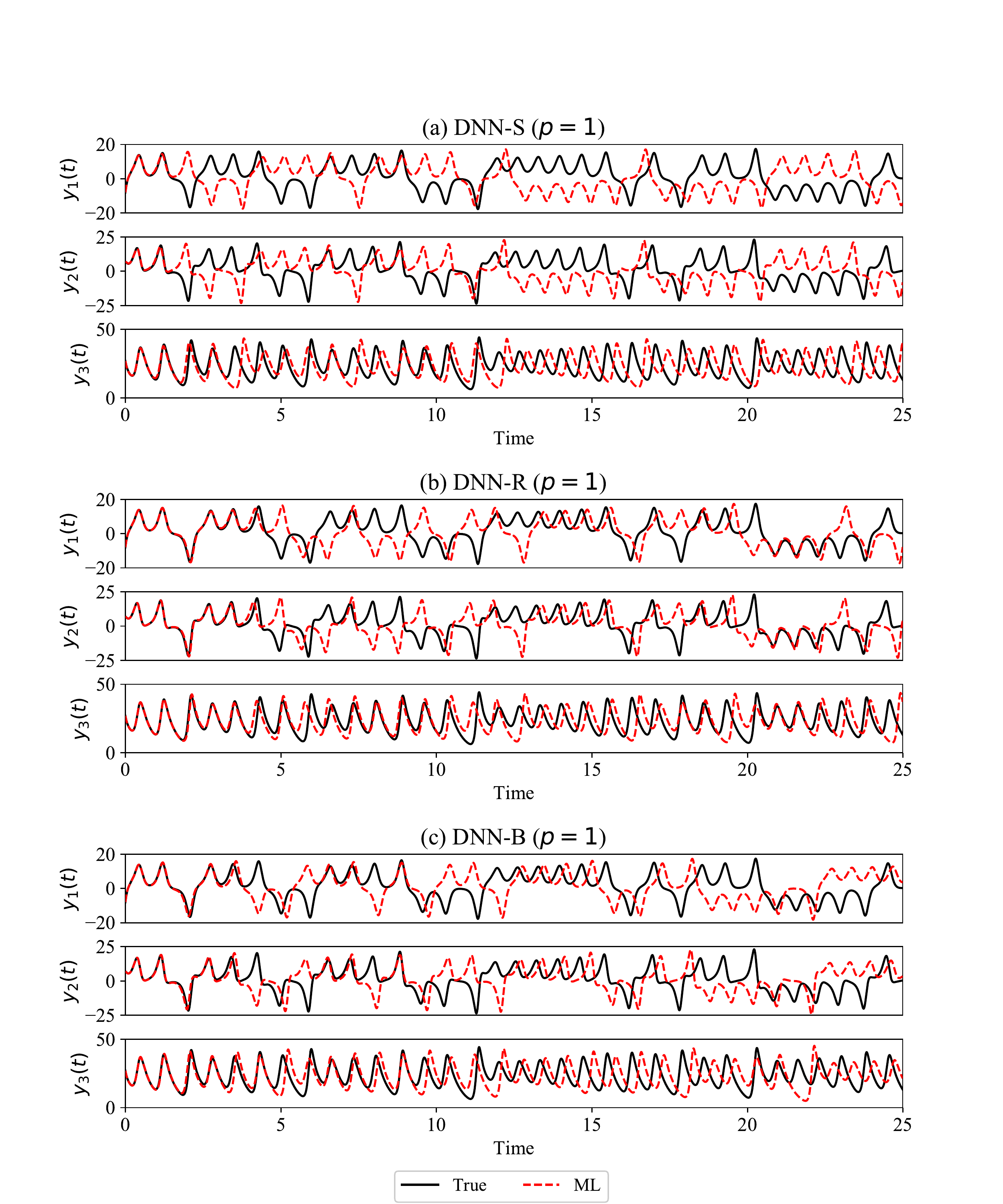}}}
\caption{Time evolution of the Lorenz system trajectories for the initial condition $[y_1 \; y_2 \; y_3]^T=[-8 \; 7 \; 27]^T$. The neural network is trained using the data generated from the true solution between $t=0$ to $25$ with $p=1$.}
\label{fig:lorenz_oneleg}
\end{figure*}

\begin{figure*}[htbp]
\centering
\mbox{
\subfigure{\includegraphics[width=0.9\textwidth]{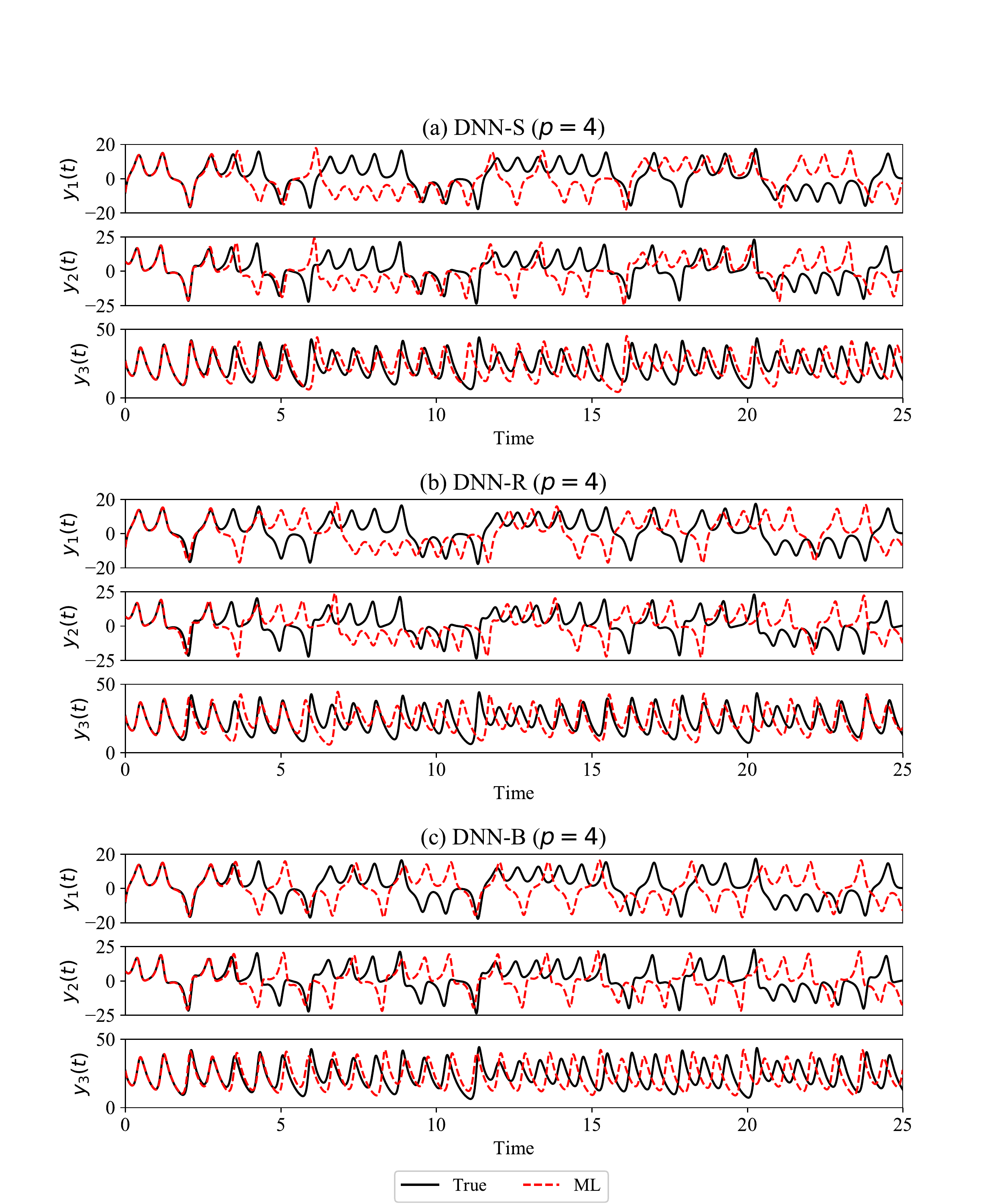}}}
\caption{Time evolution of the Lorenz system trajectories for the initial condition $[y_1 \; y_2 \; y_3]^T=[-8 \; 7 \; 27]^T$. The neural network is trained using the data generated from the true solution between $t=0$ to $25$ with $p=4$.}
\label{fig:lorenz_fourleg}
\end{figure*}

Figure \ref{fig:lorenz_oneleg} shows the true trajectory of the Lorenz system and the trajectory predicted by different DNN frameworks using $p=1$ in the input training data. The time period for which the predicted trajectory is the same as the true trajectory varies for different DNN frameworks. Lorenz system has a chaotic behavior and a smaller error in the predicted state of the system can lead to larger error in the forecasted state of the system. The time period for which the predicted trajectory follows true trajectory is longer for the DNN-R and DNN-B frameworks than the DNN-S framework. Figure \ref{fig:lorenz_fourleg} shows the similar results for $p=4$ in input training data. If we compare Figure \ref{fig:lorenz_oneleg}a and Figure \ref{fig:lorenz_fourleg}a we see that the predicted trajectory follows the true trajectory longer as we include the temporal history of the state of the system in the input to the neural network. We do not see a similar behavior for DNN-R and DNN-B frameworks. 

\begin{figure*}[htbp]
\centering
\mbox{
\subfigure{\includegraphics[width=0.8\textwidth]{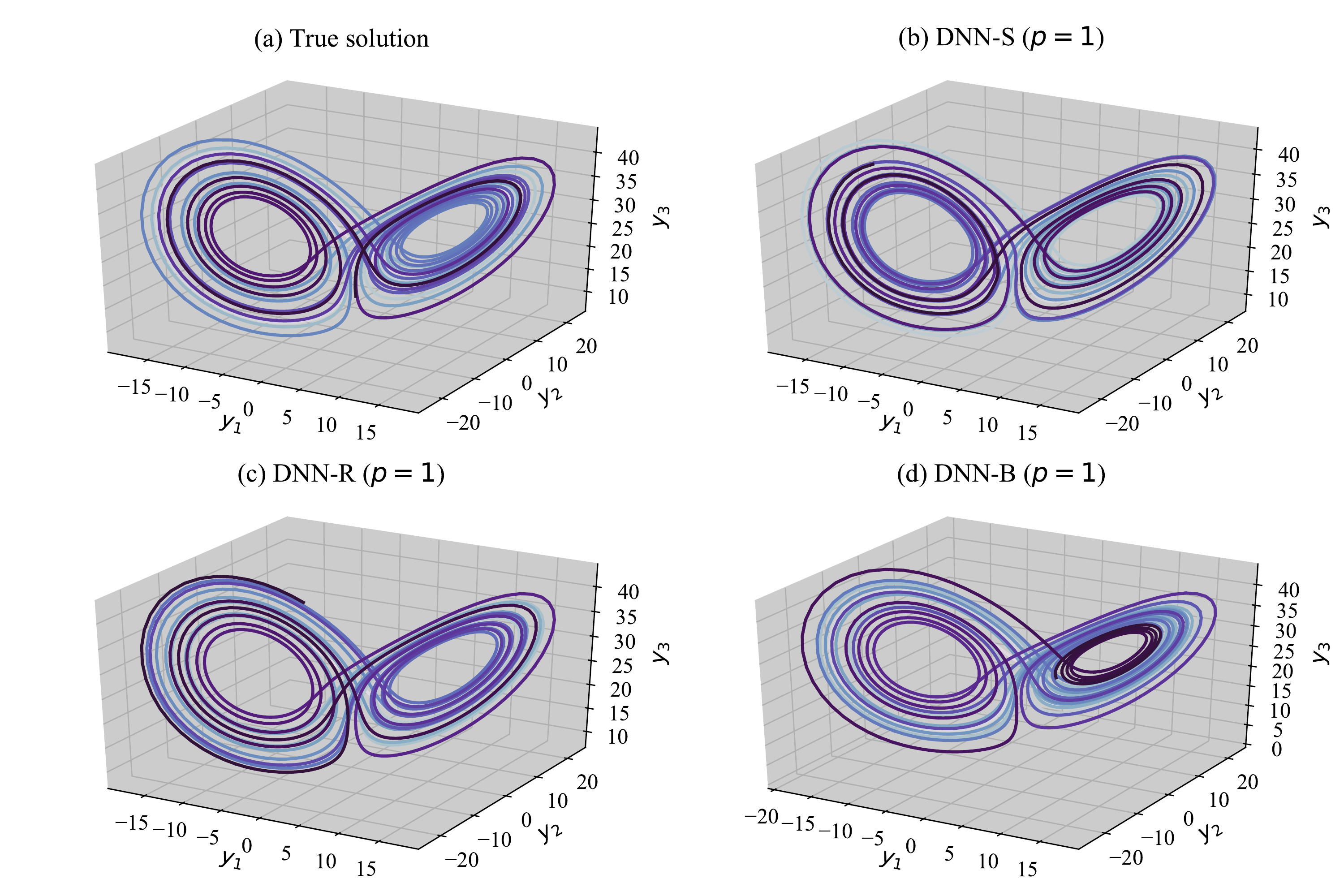}}}
\caption{The true phase portrait of the Lorenz system (a) compared with the phase portrait predicted by different DNN frameworks for time integration from $t=0$ to $t=25$. All DNN frameworks are trained with $p=1$.}
\label{fig:attractror_oneleg}
\end{figure*}

\begin{figure*}[htbp]
\centering
\mbox{
\subfigure{\includegraphics[width=0.8\textwidth]{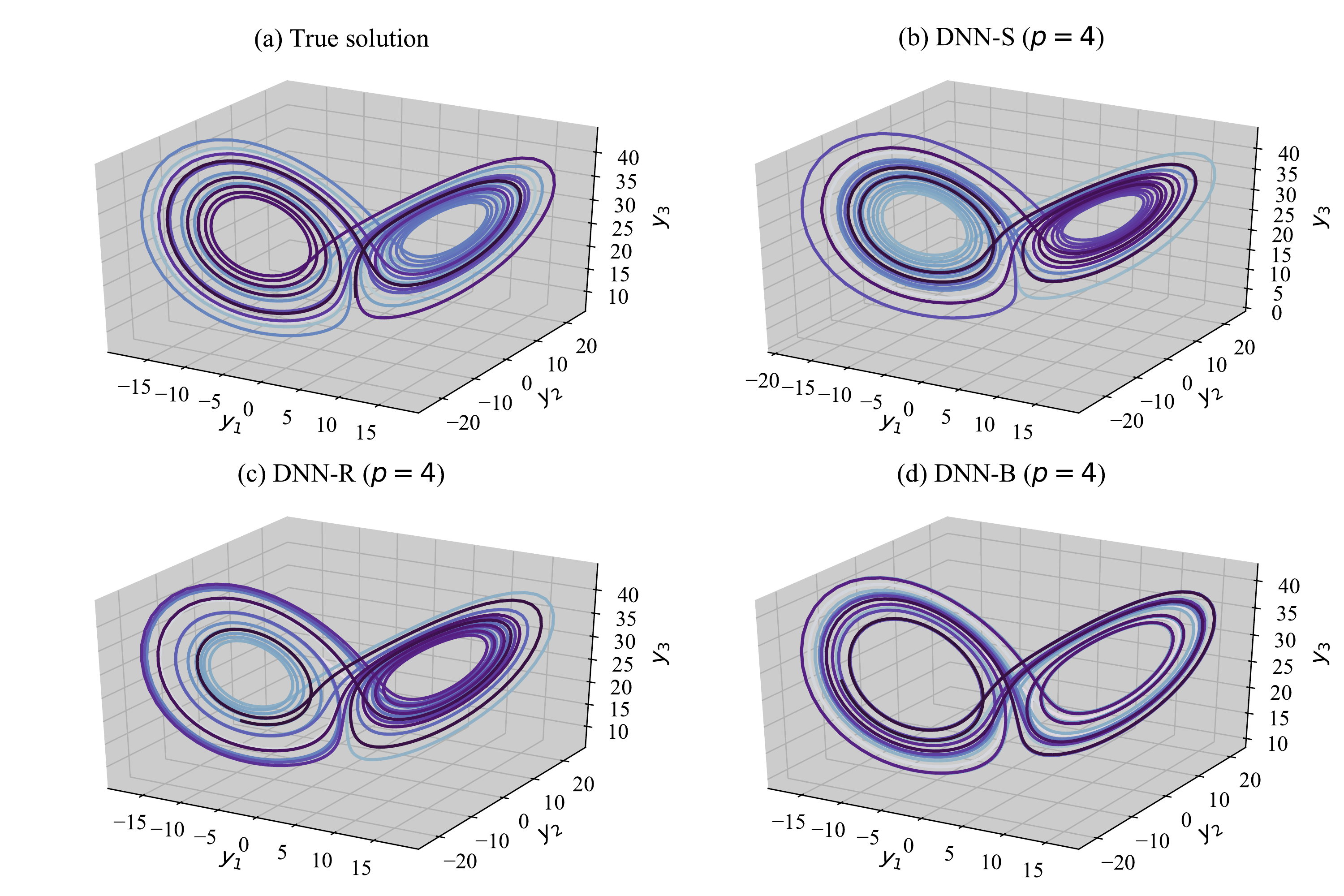}}}
\caption{The true phase portrait of the Lorenz system (a) compared with the phase portrait predicted by different DNN frameworks for time integration from $t=0$ to $t=25$. All DNN frameworks are trained with $p=4$.}
\label{fig:attractror_fourleg}
\end{figure*}

Even though the neural network is not able to predict the true trajectory of the individual state after some time, we should check the ability of the neural network to capture the overall dynamics of the Lorenz attractor. This is due to the fact that the Lorenz system has a positive Lyapunov exponent and a small perturbation in the initial condition can cause the system to diverge exponentially. Hence, a number of studies checks for the dynamics of the Lorenz attractor rather than the individual state of the system \cite{raissi2018multistep, brunton2016discovering}. Figures \ref{fig:attractror_oneleg} and \ref{fig:attractror_fourleg} show the exact dynamics of the Lorenz attractor and predicted dynamics for different DNN frameworks with $p=1$ and $p=4$, respectively. All DNN frameworks are capable of capturing the dynamics of the Lorenz system and accurately predicts the correct shape of the Lorenz attractor.          

\section{Non-intrusive Reduced Order Modeling (NIROM)} 
\label{sec:rom}
We observe a successful predictive performance for a simple nonlinear dynamical system (governed by 3 coupled ordinary differential equations) using different DNN frameworks in the previous Section. We also saw the ability of the neural network to capture the overall dynamics of the chaotic nonlinear Lorenz system. This has motivated us to test the proposed DNN frameworks for the model order reduction of a real-world test problem. 

Indeed, we transform a partial differential equation system into a set of ordinary differential equations in order to get the reduced order model. Our main motivation of this study is to convey the message that the neural network architectures can be used in developing a robust and efficient non-intrusive reduced order model. With a goal to recover the reduced order model dynamics of the underlying flow phenomena, in this Section, we implement all DNN frameworks introduced in Section \ref{sec:problem_setup} in model order reduction of a differentially heated cavity problem. This test problem setup is well-established as a model validation test case due to the problem's simplicity in terms of problem definition as well as the wide variety of applications such as nuclear reactor core isolation, solar energy storage and so on \cite{paolucci1994differentially,podvin2001low,johnston2002fourth}. At first, we will describe our non-intrusive ROM framework. Then, we will define our test problem setup along with the governing equations, then we will briefly demonstrate our non-intrusive ROM framework for the Boussinesq equation problem. Finally, we will demonstrate the comparative performance of the aforementioned DNN frameworks in terms of the temporal evolution of vorticity and temperature field and contours of temperature field for a given initial condition in Section~\ref{sec:rom_results}.

\subsection{NIROM framework}
\label{sec:nirom}
To develop our NIROM framework, we define a generalized PDE system as below
\begin{equation}
    \frac{\partial \mathbf{u}(\mathbf{x},t)}{\partial t} = \mathfrak{R}(\mathbf{u}(\mathbf{x},t); \mathbf{x},t),
\end{equation}
where $\mathbf{u}$ refers to the problems of interest (e.g., velocity, pressure, and temperature, etc) and $\mathfrak{R}$ converts the physical process possibly with linear, nonlinear, and forcing terms. 

\begin{figure*}[htbp]
\centering
\mbox{
\subfigure{\includegraphics[width=0.95\textwidth]{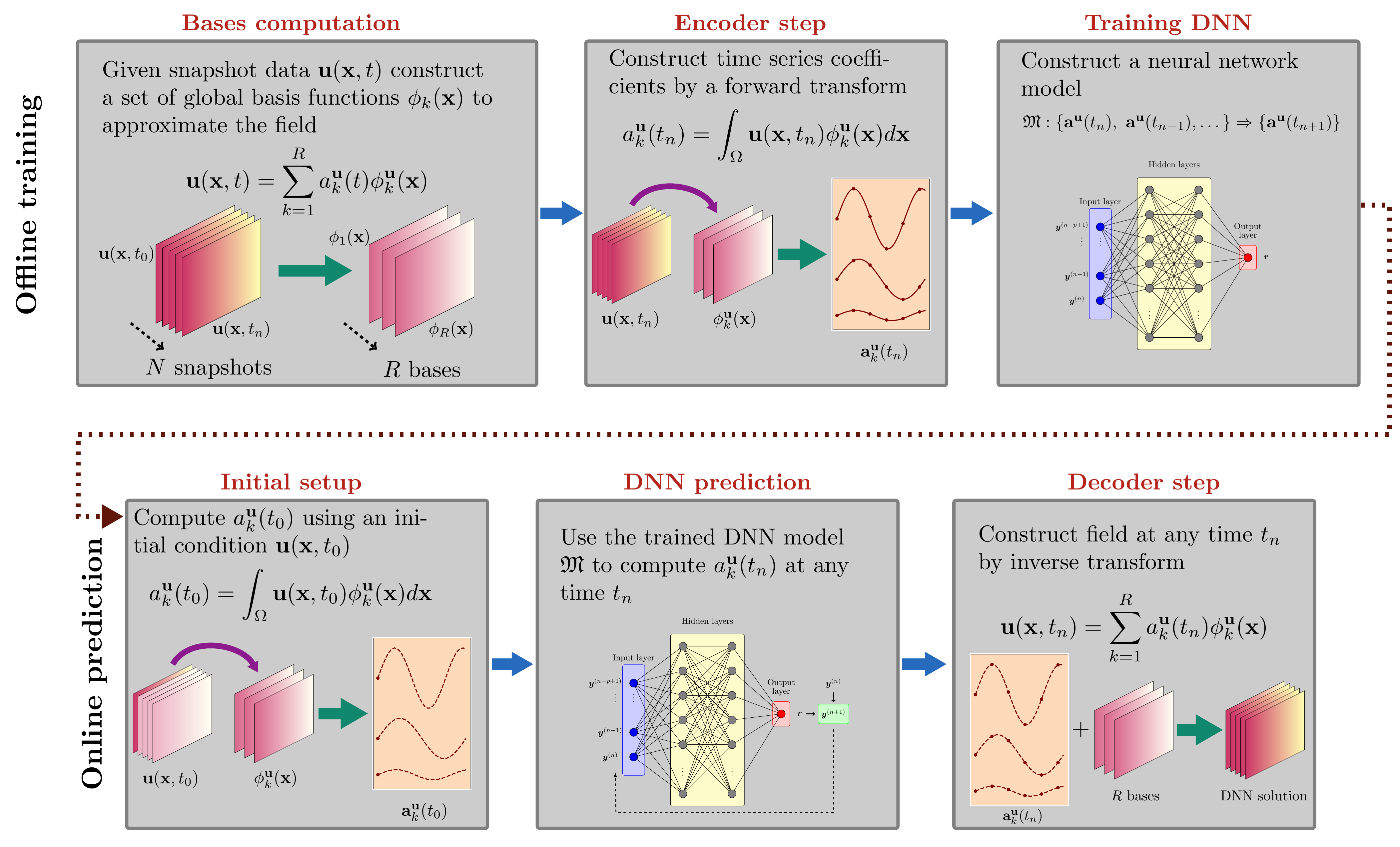}}}
\caption{Generalized non-intrusive reduced order modeling (NIROM) framework.}
\label{fig:nirom}
\end{figure*}

In our NIROM framework, we assume that we do not have access to $\mathfrak{R}(\mathbf{u}(\mathbf{x},t); \mathbf{x},t)$ operator. We formulate the proposed NIROM framework assuming that we have discrete snapshots of $\mathbf{u}(\mathbf{x},t)$. The details steps of our NIROM framework are outlined in Algorithm~\ref{alg:nirom}. We highlight that this physics-agnostic modeling approach is quite modular and decomposes the problem into the basis representation and forecasting problems. Figure~\ref{fig:nirom} depicts various stages of NIROM framework for any generalized PDE system.    

\begin{algorithm}
  \caption{NIROM framework}
  \label{alg:nirom}
  \begin{algorithmic}[1]
  
    \Statex \underline{\textit{Offline training}} \smallskip
        \State We pick or construct a set of orthonormal basis functions $\phi_k^{\mathbf{u}}(\mathbf{x})$ over domain $\Omega$ (e.g., Fourier bases, POD bases, etc)   
        \Statex \begin{equation}
            \mathbf{u}(\mathbf{x},t) = \sum_{k=1}^{R} a_{k}^{\mathbf{u}}(t)\phi_k^{\mathbf{u}}(\textbf{x}),
        \end{equation} 
        \Statex such that 
        \Statex \begin{equation}
            \int_{\Omega} \phi_i^{\mathbf{u}}(\textbf{x}) \phi_j^{\mathbf{u}}(\textbf{x}) d\mathbf{x} = \delta_{ij},
        \end{equation}
        \Statex where $\delta_{ij}$ is the Kronecker delta operator, and $\mathbf{u}(\mathbf{x},t)$ is approximated from the span of these bases  
        \State Encoder step: construct time series coefficients by a forward transform
        \Statex \begin{equation}
            a_k^{\mathbf{u}}(t_n) = \int_{\Omega} \mathbf{u}(\mathbf{x},t_n) \phi_k^{\mathbf{u}}(\textbf{x}) d\mathbf{x}.
        \end{equation}
        \State Train a time series forecasting model (e.g., a DNN model discussed in Section~\ref{sec:problem_setup})
        \Statex \begin{equation}
            \mathfrak{M}: \{\mathbf{a}^{\mathbf{u}}(t_n), \mathbf{a}^{\mathbf{u}}(t_{n-1}), \dots \mathbf{a}^{\mathbf{u}}(t_{n-p+1})\} \Rightarrow \mathbf{a}^{\mathbf{u}}(t_{n+1}).
        \end{equation}
        
    \Statex \underline{\textit{Online prediction}} \smallskip
        \State Given an initial condition $\mathbf{u}(\mathbf{x},t_0)$, compute $a_k^{\mathbf{u}}(t_0)$ using below relation
        \Statex \begin{equation}
            a_k^{\mathbf{u}}(t_0) = \int_{\Omega} \mathbf{u}(\mathbf{x},t_0) \phi_k^{\mathbf{u}}(\textbf{x}) d\mathbf{x}.
        \end{equation}
        \State Use the trained DNN model $\mathfrak{M}$ to predict $a_k^{\mathbf{u}}(t_n)$ at any time $t_n$
        \State Decoder step: construct the field at any time $t_n$ by inverse transform
        \Statex \begin{equation}
            \mathbf{u}(\mathbf{x},t_n) = \sum_{k=1}^{R} a_{k}^{\mathbf{u}}(t_n)\phi_k^{\mathbf{u}}(\textbf{x}).
        \end{equation}
  \end{algorithmic}
\end{algorithm}

\subsection{Problem definition: Boussinesq equations}
\label{subsec:boussinesq}
For this numerical experiment, we investigate the convective flow behavior in a two-dimensional buoyancy-driven flow in a differentially heated tall cavity \cite{liu2003fourth}. We consider the following dimensionless form of the two-dimensional incompressible Boussinesq equations \cite{san2015principal,deane1991computational,sirovich1991computational,san2018machine} on our computational domain, $\Omega$:
\begin{align}
\label{eq:ge}
&\nabla \cdot {\bf u} = 0, \\
&\frac{\partial {\bf u}}{\partial t} + ({\bf u}\cdot\nabla) {\bf u} =
-\nabla p + \frac{1}{\text{Re}} \nabla^2 {\bf u} + \text{Ri} \ \theta \ \hat{e}_{j}, \\
&\frac{\partial \theta}{\partial t} + ({\bf u}\cdot\nabla) \theta =
\frac{1}{\text{Re} \ \text{Pr}} \nabla^2 \theta,
\end{align}
where $\bf u$ refers to the velocity vector, $p$ and $\theta$ denote the pressure and temperature fields, respectively, and $\hat{e}_{j}$ is the unit vector in $y$ direction. Here, $\nabla$ and $\nabla^2$ are the standard two-dimensional differential and Laplacian operator, respectively. In general, the Boussinesq equations are characterized by three dimensionless numbers: Reynolds number (Re), Prandtl number (Pr) and Richardson number (Ri). However, other relevant dimensionless numbers can be introduced into the equation as a control parameter based on the physics of the system. For example, we introduce Rayleigh number (Ra) in our study due to the natural convection heat transfer which can be expressed as

\begin{align}
\label{eq:ra}
\text{Ra} = \text{Ri} \ \text{Re}^2 \ \text{Pr}.
\end{align}
In our problem setup, we fix Pr$= 0.71$, Ri $=1$. In our two-dimensional full order model (FOM) simulation, we utilize the vorticity-streamfunction formulation to avoid numerical complexity associated with the primitive variable formulation \cite{san2018machine}. Therefore, we introduce the vorticity ($\omega =\nabla \times \bf u$) and streamfunction ($\psi$) for Equation~(\ref{eq:ge}) specified by the following coupled equations \cite{johnston2002fourth,san2015principal}:
\begin{align}\label{eq:vs1}
  & \frac{\partial \omega}{\partial t}  + \text{J}(\omega, \psi) = \frac{1}{\text{Re}}\nabla^2 \omega + \text{Ri} \frac{\partial \theta}{\partial x}, \\ \label{eq:vs2}
  & \frac{\partial \theta}{\partial t}  + \text{J}(\theta, \psi) = \frac{1}{\text{Re} \ \text{Pr}}\nabla^2 \theta,
\end{align}
where Jacobian J accounts for the nonlinear advection term, which is defined as
\begin{align}
  \text{J}(f, g) = \frac{\partial g}{\partial y}\frac{\partial f}{\partial x} - \frac{\partial g}{\partial x}\frac{\partial f}{\partial y}.
\end{align}
The flow velocity components can be found from the stream function, $\psi$, using following definitions:
\begin{align}\label{sf}
  u = \frac{\partial \psi}{\partial y}, \ \ v = - \frac{\partial \psi}{\partial x}.
\end{align}
The kinematic equation connecting the vorticity and streamfunction can be found by substituting the velocity components in terms of stream function, which form the following divergence-free constraint satisfying the Poisson equation:
\begin{align}\label{pois}
  \nabla^2 \psi = - \omega.
\end{align}
Our Cartesian computational domain is $(x,y) \in [0,\ 1]\times[0,\ 8]$. We utilize wall boundary conditions for all four sides of our computational domain with an adiabatic condition on the top and bottom wall. We imply Dirichlet conditions on left ($\theta = 0.5$) and right ($\theta = - 0.5$) walls. At the cavity walls, we enforce no-slip boundary conditions by setting zero values for streamfunction and vorticity values calculated by the Briley's formula \cite{maulik2017dynamic}. A detailed derivation and discussion on our test problem setup along with the numerical schemes for generating snapshots through FOM simulation can be found in a recent study conducted by San and Maulik \cite{san2018machine}. Note that, even though the simulation is performed for a maximum time of $t = 1000$, we perform all our statistical analysis from $t = 900$ (after an initial transient period) at $128 \times 1024$ grid resolutions, and collect snapshot data only between $t=900$ and $t=950$, and perform our quantitative analyses for the prediction of both in-sample data zone (between $t=900$ and $t=950$) as well as the out-of-sample data zone (between $t=950$ and $t=1000$). For the rest of this paper, we shall refer to this initial state (i.e., time $t=900$ in our physical simulation) as $t=0$ for prescribing initial conditions for ROMs. Therefore, our in-sample data zone spans between $t=0$ and $t=50$, and our out-of-sample data zone spans between $t=50$ and $t=100$. We use the data in the in-sample-zone for training the neural network.  

\begin{figure*}[htbp]
\centering
\mbox{
\subfigure{\includegraphics[width=0.95\textwidth]{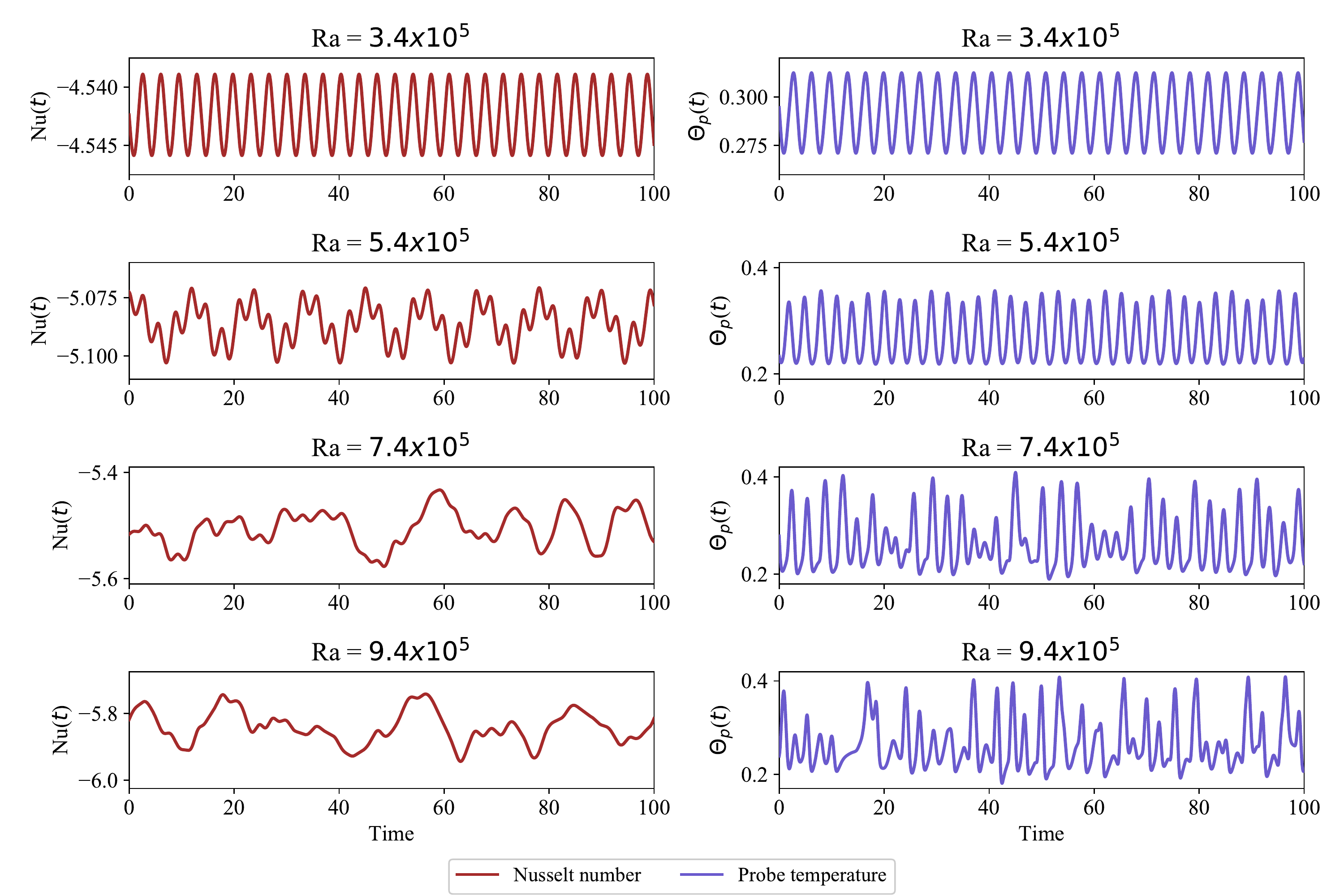}}}
\caption{Evolution of the Nusselt number Nu$(t)$ at $x=0$ (left) and designated probe temperature $\theta_p$. The temperature $\theta_p$ is probed at the location $x=0.125$ and $y=7.0$.}
\label{fig:ra_effect}
\end{figure*}

The DNS technique used for performing the FOM simulation is validated by recording the simulation statistics at a designed probe point and comparing it with the studies in the literature \cite{tyliszczak2016high}. We simulate the differentially heated cavity problem for four different Rayleigh numbers. The flow is smooth and orderly for lower Rayleigh numbers. As we increase the Rayleigh number, the flow starts getting chaotic and turbulent. We calculate the Nusselt number along the vertical left wall ($x=0$) using the below formula
\begin{equation}
    \text{Nu($t$)} = \frac{1}{H} \int_{0}^{H} \frac{\partial \theta}{\partial x} \Big|_{x=0} dy,
    \label{eq:nusselt}
\end{equation}
where $H=8$. Figure \ref{fig:ra_effect} shows the statistics of Nusselt number along the left wall and the designated probe temperature ($x=0.125$ and $y=7.0$). It can be seen that the flow behavior is periodic for both Nusselt number and the temperature history at the probed location at lower Rayleigh numbers. As the Rayleigh number increases, we see that the Nusselt number and probed temperature variation is not periodic due to the turbulent nature of flow. We test our non-intrusive framework for Ra = $3.4\times10^5$ where the flow is periodic and for Ra = $9.4\times10^5$ where the flow is chaotic.       

\subsection{NIROM framework for Boussinesq equations}  
\label{subsec:ni_rom_boussinesq}

\begin{figure*}[htbp]
\centering
\mbox{
\subfigure{\includegraphics[width=0.95\textwidth]{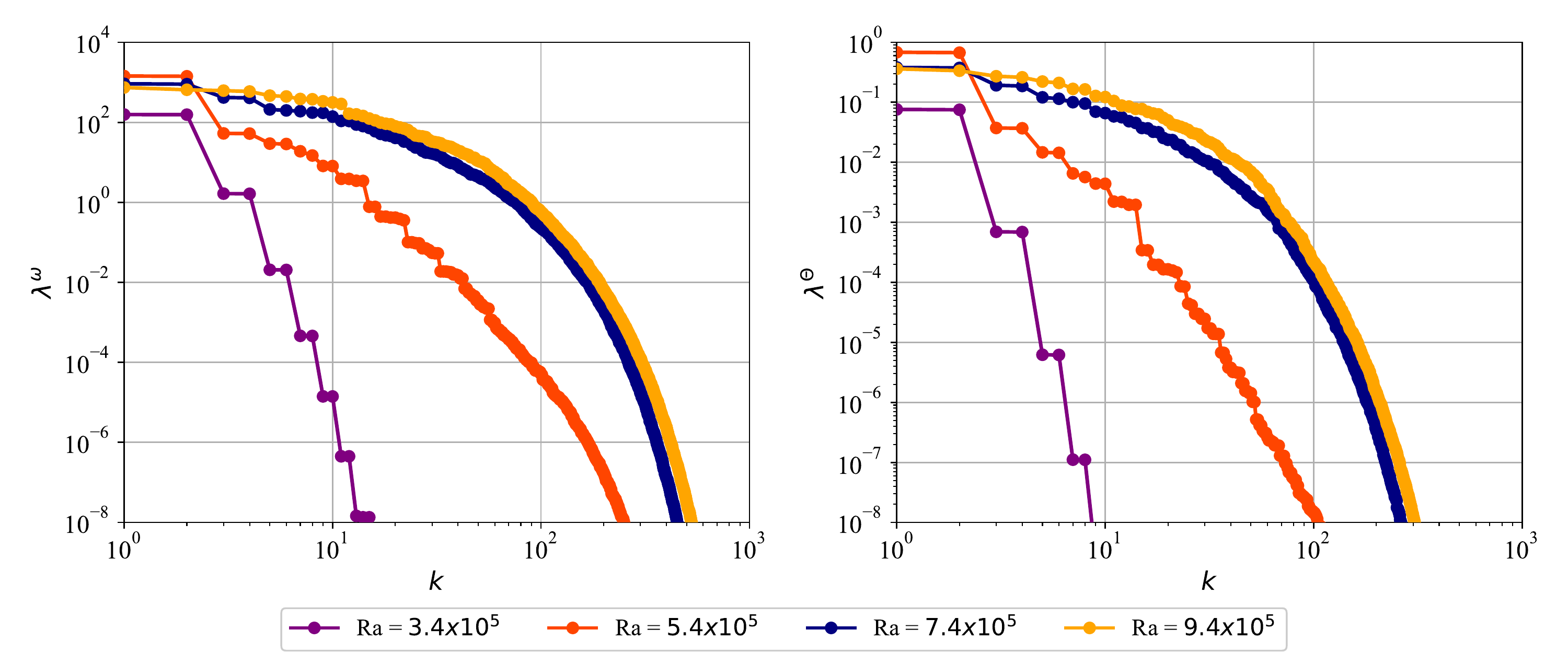}}}
\caption{Eigenvalues of the correlation matrix $C$ using $M=1000$ snapshots for different Rayleigh numbers.}
\label{fig:eigen}
\end{figure*}

In this Section, we present our non-intrusive ROM setup for the unsteady, incompressible Boussinesq equations given by Equation~(\ref{eq:vs1}) and Equation~(\ref{eq:vs2}). In our ROM settings, we first compute the desired set of orthogonal spatial basis functions from a stored high-fidelity data snapshots using proper orthogonal decomposition (POD). We obtain the data snapshots from a high-resolution FOM simulation. Using the pre-computed basis functions, we develop the non-intrusive framework in an encoder-decoder approach using different DNN frameworks proposed in Section \ref{sec:problem_setup}. To compare the predictive performance of the non-intrusive ROMs with respect to the standard intrusive ROM, we develop our intrusive ROM framework (ROM-G) using the Galerkin projection to derive the dynamical model for the POD coefficients \cite{rowley2004model,lucia2004reduced,iollo2000stability}. The implementation of Galerkin projection for the underlying test problem is detailed in the recent work by San and Maulik \cite{san2018machine} and hence, is not discussed in the present work. Here, we demonstrate the development of the non-intrusive ROM methodologies briefly before proceeding to the numerical results for analyses.

The POD bases for the vorticity field can be constructed from the field variable $\omega(x, y)$ at different time steps which we denote as snapshots, i.e., for M number of snapshots, $\omega(\textbf{x}, t_n)$ are the stored snapshots for $n=1,2,\dots,M$. The time-averaged field can be computed as
\begin{align}\label{POD_1}
  \bar{\omega}(\textbf{x}) = \frac{1}{M} \sum_{n=1}^{M} \omega(\textbf{x},t_n).
\end{align}
To map the snapshot data to its origin, we then compute the mean-subtracted snapshots or the fluctuating fields by
\begin{equation}
\omega'(\textbf{x},t_n)=\omega(\textbf{x},t_n)-\bar{\omega}(\textbf{x}).
\end{equation}
This subtraction guarantees that ROM solution would satisfy the same boundary conditions as the full order model \cite{chen2012variants}. To simplify the eigenvalue problem necessary for POD bases calculation, we utilize the standard method of snapshots proposed by Sirovich \cite{sirovich1987turbulence} which reduces the larger dimension problem to a much smaller dimension problem. To do so, we construct a $M\times M$ correlation matrix of the fluctuating part ${C}=[c_{ij}]$ which is computed from the inner product of the mean-subtracted snapshots
\begin{equation}
c_{ij}=\langle \omega'(\textbf{x},t_i), \omega'(\textbf{x},t_j)\rangle,
\end{equation}
where $i$ and $j$ refer to the snapshot indices. The definition of the inner product of any arbitrary two fields $f$ and $g$ can be expressed as
\begin{align}\label{POD_3}
  \langle f,g\rangle = \int_{\Omega} f(\textbf{x}) g(\textbf{x}) d\textbf{x}.
\end{align}
In the current study, we use the well-known Simpson's integration rule for a numerical computation of the inner products. Next, an eigendecomposition of $C$ matrix is performed by solving
\begin{align}
\label{POD_4}
    CW=W\mathbf{\Lambda} \, ,
\end{align}
where $\mathbf{\Lambda}$ is a diagonal matrix whose entries are the eigenvalues $\lambda_k^{\omega}$ of $C$, and $W$ is a matrix whose columns $w_k$ are the corresponding eigenvectors. This has been shown in detail in various POD literature (see, e.g., \cite{sirovich1987turbulence,ravindran2000reduced}). It should be noted that eigenvalues need to be arranged in a descending order (i.e., $\lambda_1^{\omega}\ge\lambda_2^{\omega}\ge\dots\ge\lambda_M^{\omega}$), for proper selection of the POD modes. The POD modes of vorticity field $\phi_{k}^{\omega}$ are then computed as
\begin{equation}
\phi_{k}^{\omega}(\textbf{x})=\dfrac{1}{\sqrt{\lambda_k^{\omega}}}\sum_{n=1}^{M} w^{n}_{k} \omega'(\textbf{x},t_n),
\end{equation}
where $w^{n}_{k}$ is the $n$th component of the eigenvector $W$. The scaling factor, $\left(\dfrac{1}{\sqrt{\lambda_k^{\omega}}}\right)$ is to guarantee the orthonormality of POD modes i.e., $\langle \phi_i^{\omega}, \phi_j^{\omega}\rangle = \delta_{ij}$. Here, $\delta_{ij}$ is the Kronecker delta defined by
\begin{equation}\label{eq:inner2}
 \delta_{ij} = \begin{cases} 1, & \text{if } i=j,\\ 0, & \text{if } i\neq j. \end{cases}
\end{equation}
Similarly, we can compute the POD bases for the temperature field which is $\phi_{k}^{\theta}(\textbf{x})$.

\begin{figure*}[htbp]
\centering
\mbox{
\subfigure{\includegraphics[width=0.9\textwidth]{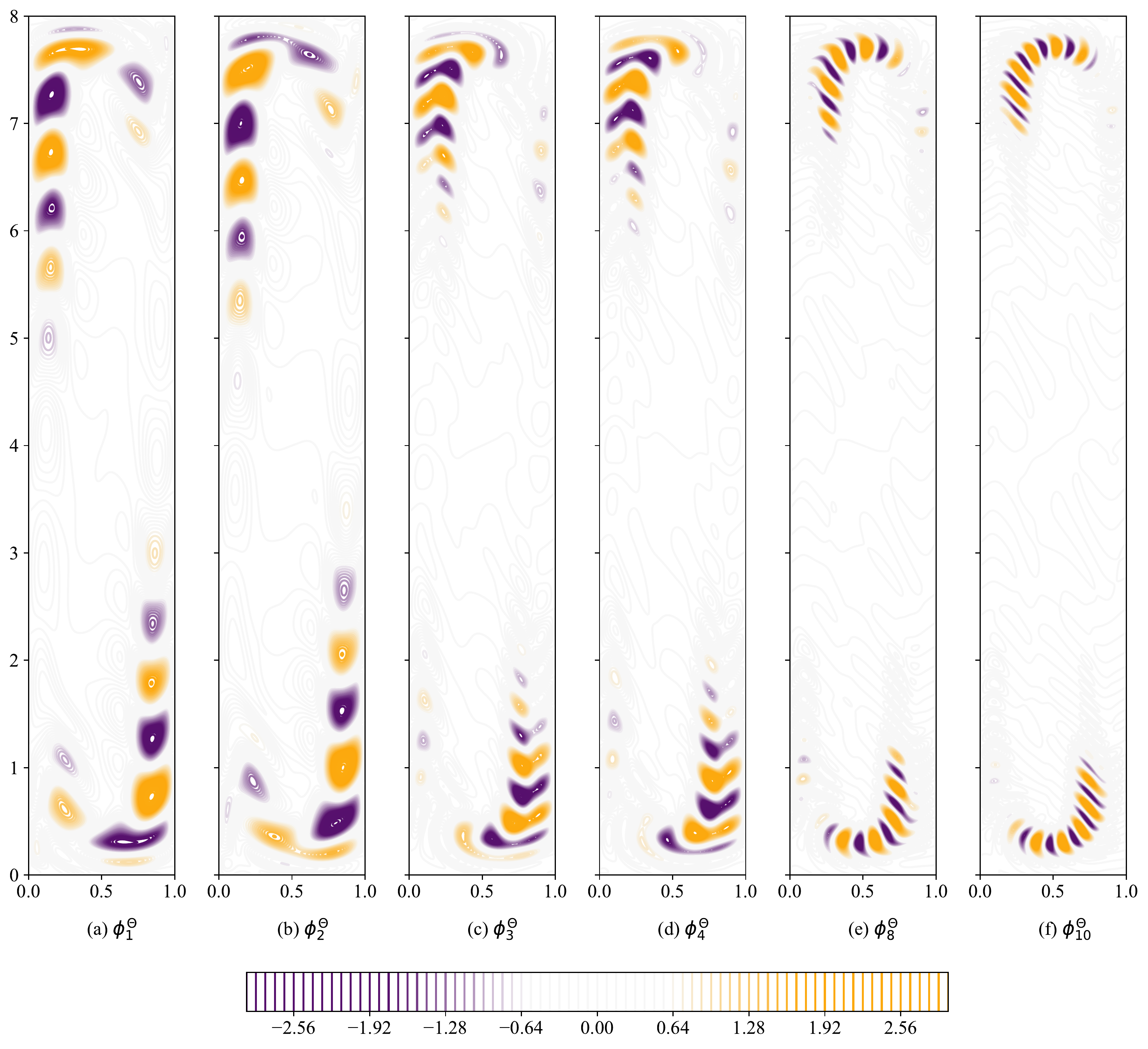}}}
\caption{Illustrative contour plots for some of the POD basis functions for temperature field at Ra = $3.4\times10^5$ for the differentially heated cavity problem.}
\label{fig:basis34}
\end{figure*}

\begin{figure*}[htbp]
\centering
\mbox{
\subfigure{\includegraphics[width=0.9\textwidth]{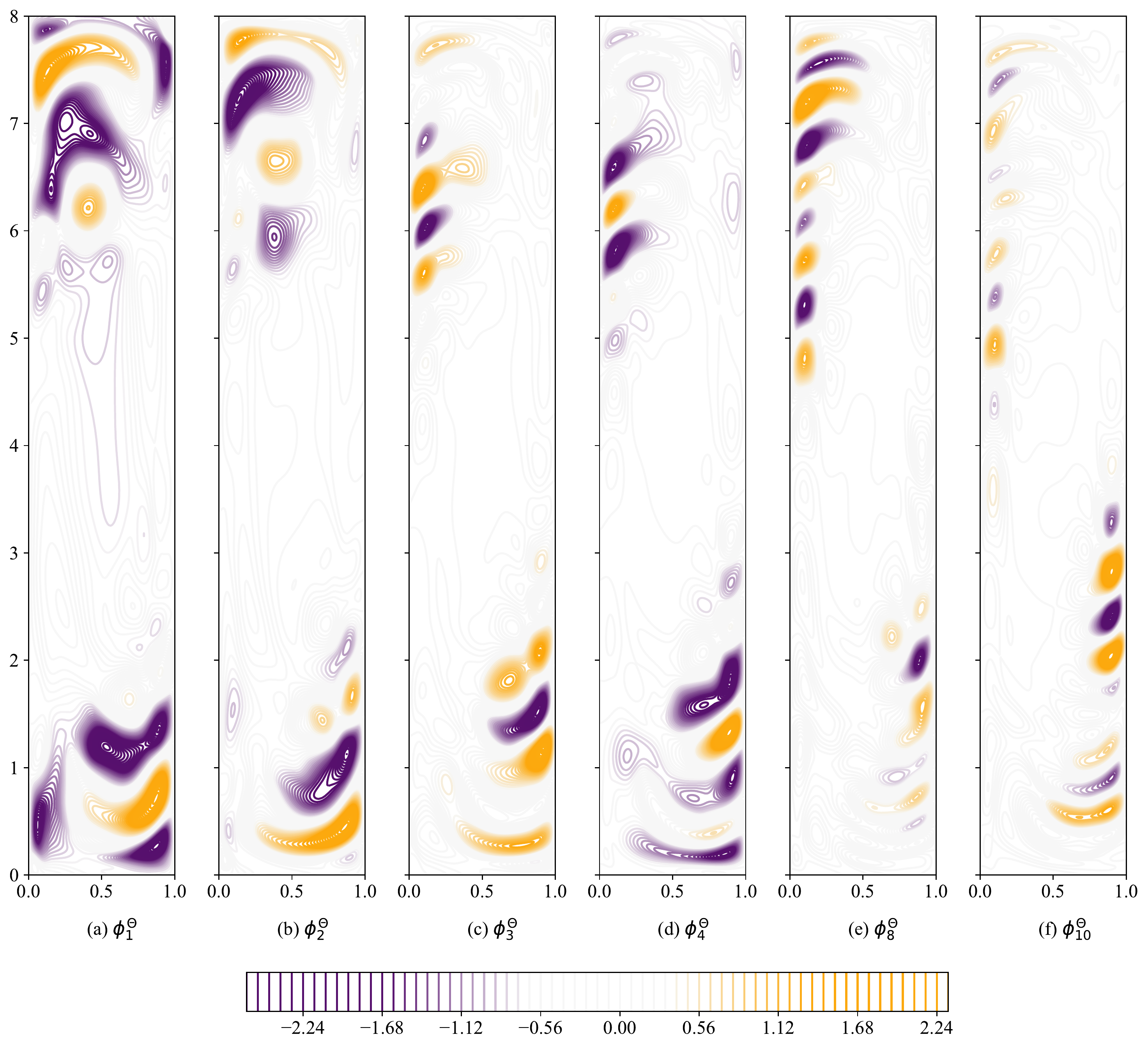}}}
\caption{Illustrative contour plots for some of the POD basis functions for temperature field at Ra = $9.4\times10^5$ for the differentially heated cavity problem.}
\label{fig:basis94}
\end{figure*}

Figure \ref{fig:eigen} shows the decay of eigenvalues of the correlation matrix for vorticity and temperature field. We can observe that the first 10 modes are able to capture more than 99\% of the total energy for both vorticity and temperature at lower Rayleigh numbers (Ra = $3.4\times10^5$ and Ra = $5.4\times10^5$). Also, there is a faster decay of eigenvalues for lower Rayleigh numbers than for higher Rayleigh numbers. For higher Rayleigh numbers, the first 10 modes are not sufficient to capture a major portion of the energy. For example, the first 10 modes of vorticity field capture only 76\% and 69\% of the total energy for vorticity at Rayleigh numbers Ra = $7.4\times10^5$ and Ra = $9.4\times10^5$, respectively. In order to capture more than 95\% of the energy, we will need to consider more than 40 modes. If we include 40 modes, then the Galerkin projection becomes computationally expensive and we lose the benefit of ROM framework. If we include only 10 modes, then the Galerkin projection is unbounded, as we will see in Section \ref{sec:rom_results}, and it gives a physically wrong solution. We demonstrate that our non-intrusive framework give sufficiently accurate results comparable to the true projection of the FOM solution on reduced order space even with less number of POD modes. It is evident that if we include a higher number of modes to capture the total energy, the true projection of FOM solution on lower dimensional bases will approximate the FOM solution. However, the computational burden will also go up with an increased number of modes. In Figure \ref{fig:basis34}, we provide the contour plots for a few POD basis for the temperature field $\theta$ to indicate the structure of the solution at Ra = $3.4\times10^5$. We can observe that the solution is smooth and periodic for lower Rayleigh number case and only small structures are truncated after 10 POD modes. On the other hand, we observe from Figure \ref{fig:basis94}, that there are still some of the large structures remaining in the 10th mode for higher Rayleigh number case. This means that some of the important flow features are truncated due to consideration of only the first 10 dominant POD modes.   

To obtain the non-intrusive ROM, we utilize an encoder-decoder approach to transfer data from full order space to reduced order space and vice versa. During the encoder stage, we transform data from the full order space to the reduced order space by using the following projection for both field parameters:
\begin{align}
a_k (t) &= \langle \omega(\textbf{x}, t) -  \bar{\omega}(\textbf{x}), \phi_k^{\omega}(\textbf{x}) \rangle, \\
b_k (t) &= \langle \theta(\textbf{x}, t) -  \bar{\theta}(\textbf{x}), \phi_k^{\theta}(\textbf{x}) \rangle.
\end{align}
So, at initial time $t = 0$, we can compute the initial conditions by
\begin{align}\label{eq:ic1}
a_k(0) &= \big\langle \omega(\textbf{x},0) - \bar{\omega}(\textbf{x}), \phi_{k}^{\omega}(\textbf{x}) \big\rangle, \\
b_k(0) &= \big\langle \theta(\textbf{x},0) - \bar{\theta}(\textbf{x}), \phi_{k}^{\theta}(\textbf{x}) \big\rangle,
\end{align}
where $\omega(\textbf{x},0)$ and $\theta(\textbf{x},0)$ is the vorticity and temperature field, respectively, specified at initial time. Here, $a_k (t)$ and $b_k (t)$ are the time-dependent modal coefficients for the vorticity and temperature, respectively. This will form an initial value two equations ODE system similar to the problem in Section~\ref{sec:dynsystem} where we require to solve $d a_k/d t$ and $d b_k/d t$ until the final time. As discussed earlier, we utilize the different DNN frameworks and time histories as input to predict the temporal evolution of $a_k (t)$ and $b_k (t)$. For the prediction with sequential time history data, we predict the next time step using the DNN-S framework with one ($p=1$) and four ($p=4$) legs temporal history in the input. We also employ DNN-R and DNN-B frameworks which predicts the residual and the numerical slope, respectively based on one ($p=1$) and four ($p=4$) legs temporal history in the input to the neural network. In the decoder stage, we reconstruct the reduced order solution to the full order solution by using the following definition:
\begin{align}
    \omega(\textbf{x}, t) &= \bar{\omega}(\textbf{x}) + \sum_{k=1}^{R} a_{k}(t)\phi_k^{\omega}(\textbf{x}), \\
    \theta(\textbf{x}, t) &= \bar{\theta}(\textbf{x}) + \sum_{k=1}^{R} b_{k}(t)\phi_k^{\theta}(\textbf{x}), \label{eq:deco}
\end{align}
where $R \ (<< M)$ is the retained most energetic POD modes. Since we do not use any physical equations for the time integration of the solution field, we can say our ROM setup using DNN is fully non-intrusive. 

\section{Numerical results for Boussinesq equations}
\label{sec:rom_results}
To evaluate the performance of our DNN frameworks within the underlying non-intrusive ROM setup, we present the time series evolution for the vorticity transport equation. In addition, we compare the temperature field at the final time (i.e., $t=100$) predicted using non-intrusive ROM framework, with FOM simulation and its projection on reduced order space (True projection). We further evaluate the performance of DNN frameworks in predicting engineering quantities of interest such as time-averaged Nusselt number. We also perform the quantitative assessments of different model's predictive performance in terms of the RMSE calculated using Equation~(\ref{eq:rmse}). We present our analysis for two Rayleigh numbers: Ra = $3.4\times10^5$ where the flow is smooth and periodic, and Ra = $9.4\times10^5$ where the flow is turbulent and chaotic.    
\begin{figure*}[htbp]
\centering
\mbox{
\subfigure{\includegraphics[width=0.9\textwidth]{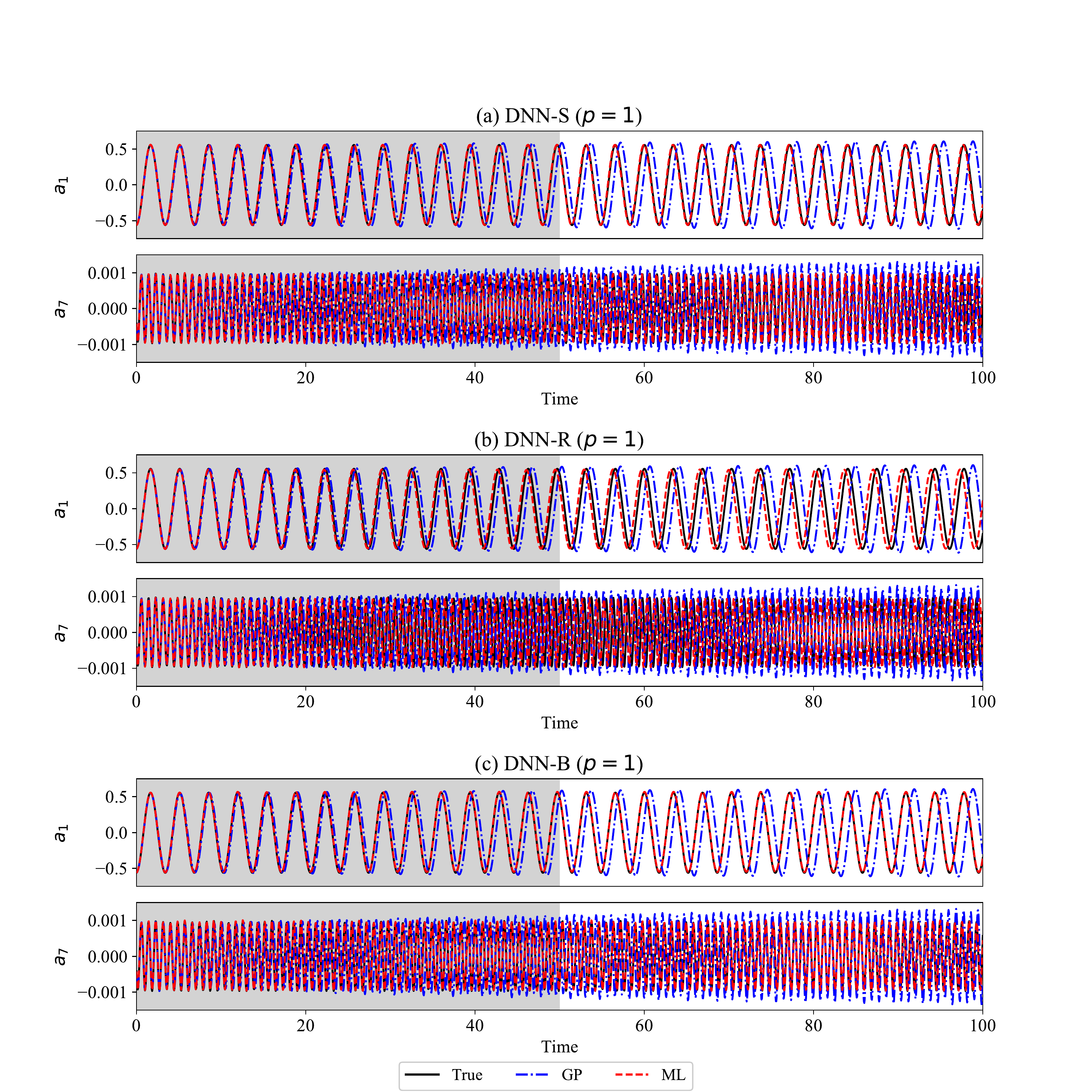}}}
\caption{Evolution of temporal coefficients for vorticity transport equation at Ra = $3.4\times10^5$ for different frameworks with $p=1$. The neural network is trained using the data highlighted in light gray color in the above figure.}
\label{fig:a34_oneleg}
\end{figure*}

\begin{figure*}[htbp]
\centering
\mbox{
\subfigure{\includegraphics[width=0.9\textwidth]{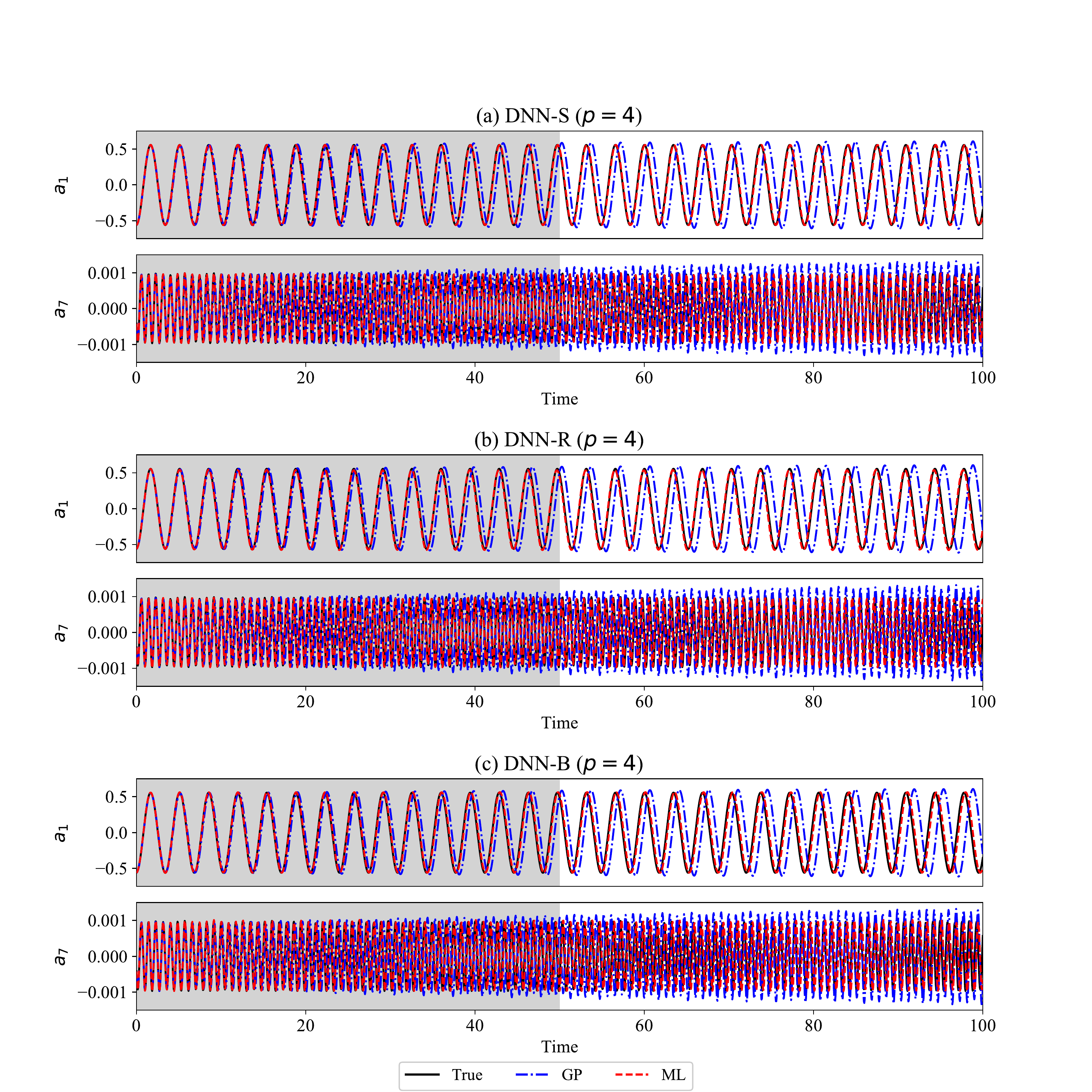}}}
\caption{Evolution of temporal coefficients for vorticity transport equation at Ra = $3.4\times10^5$ for different frameworks with $p=4$. The neural network is trained using the data highlighted in light gray color  in the above figure.}
\label{fig:a34_fourleg}
\end{figure*}

For all DNN frameworks, we use six hidden layers with 120 neurons each. The maximum number of iterations is set to 900 and 10\% of the data is used for validation to avoid overfitting. The root mean square error used for evaluating the quantitative performance of DNN frameworks measures the difference between the true data and the predicted data for each mode.  Hence, we present the true and predicted trajectories by neural network for only two modes $a_1$ and $a_7$ for conciseness. Additionally, we compare the modal coefficient trajectory with the POD Galerkin projection (GP) trajectories (i.e., time dependent amplitude coefficients of ROM-G). \textcolor{rev1}{
It should be noted that our criteria for selection of hyperparameters are not necessarily optimal but are based on heuristics (involving different activation functions, number of layers/neurons/iterations etc.) that enable our neural networks to accurately predict time evolution of dynamical systems. We also highlight that there are statistical methods available to select hyperparameters such as Bayesian optimization\cite{snoek2012practical}.}      

As illustrated in Figure \ref{fig:a34_oneleg} and for the rest of our analysis in this Section, we display the true solution as a black solid line, and regular POD-Galerkin projection based ROM, i.e., ROM-G solution as a blue dash-dot line. The solution predicted by the neural network is shown by the dashed red line. The training data for the neural network is taken from the time series of modal coefficients obtained by projecting the FOM solution on the reduced order space between $t=0$ to $t=50$. This is consistent with the snapshot data used for generating the POD bases. The training data for the neural network is highlighted using the light gray color in all time series plots. When we test the neural network, we start with an initial condition at $t=0$ and proceed in an iterative fashion as discussed in Section \ref{sec:problem_setup}. Therefore the data between $t=0$ to $t=50$ is in-sample data and $t=50$ to $t=100$ is the out-of-sample data. The neural network has seen the in-sample data during training and hence is expected to give a good prediction for that time period. The question to ask is how does a neural network predict for the out-of-sample data.

Figure~\ref{fig:a34_oneleg} shows the time evolution of two modal coefficients of vorticity transport equation for all DNN frameworks using $p=1$ in the input training data for Ra = $3.4\times10^5$. It can be clearly seen that there is a phase difference between the true projection and Galerkin projection for the first modal coefficient $a_1$. Also, the Galerkin projection predicts slight amplification in the amplitude of $a_1$, especially near the final time. The similar amplification in magnitude is also observed for $a_7$ with the Galerkin projection. The prediction by all DNN frameworks is better than the Galerkin projection and the prediction is very close to the true projection of the FOM solution on reduced order space. For the DNN-R framework, there is a slight phase shift in DNN prediction with respect to the true projection as the time proceeds. The prediction for the DNN-R framework can be improved by including the past history of the modal coefficients in the input training data. Figure~\ref{fig:a34_fourleg} shows the DNN prediction with $p=4$ in the input training data. We can see that the prediction is improved for the DNN-R framework using $p=4$. The prediction for DNN-S and DNN-B frameworks was already good using $p=1$ and remains the same with $p=4$ also. We see a similar prediction for other modal coefficients also.

\begin{figure*}[htbp]
\centering
\mbox{
\subfigure{\includegraphics[width=0.9\textwidth]{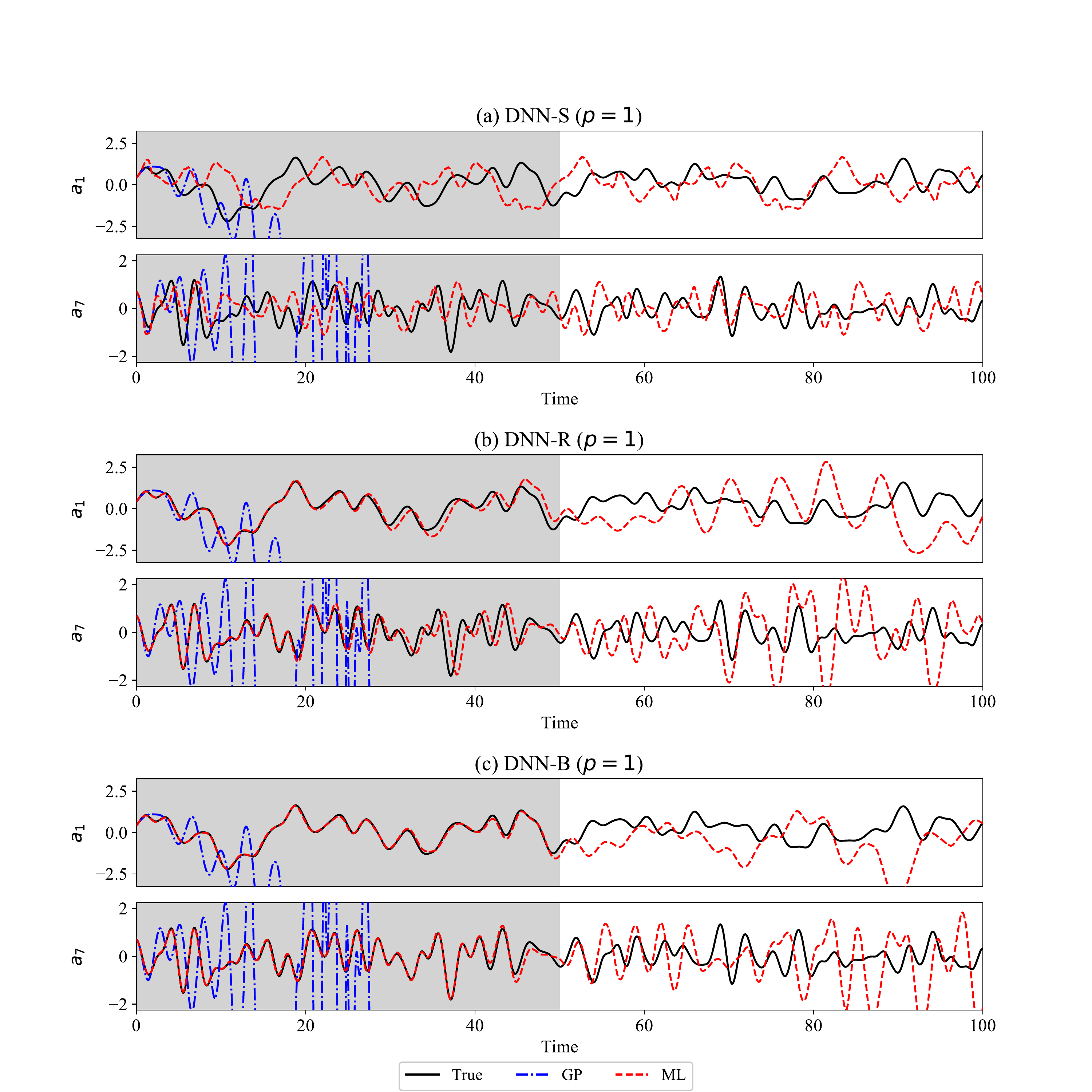}}}
\caption{Evolution of temporal coefficients for vorticity transport equation at Ra = $9.4\times10^5$ for different frameworks with $p=1$. The neural network is trained using the data highlighted in light gray color in the above figure.}
\label{fig:a94_oneleg}
\end{figure*}

\begin{figure*}[htbp]
\centering
\mbox{
\subfigure{\includegraphics[width=0.9\textwidth]{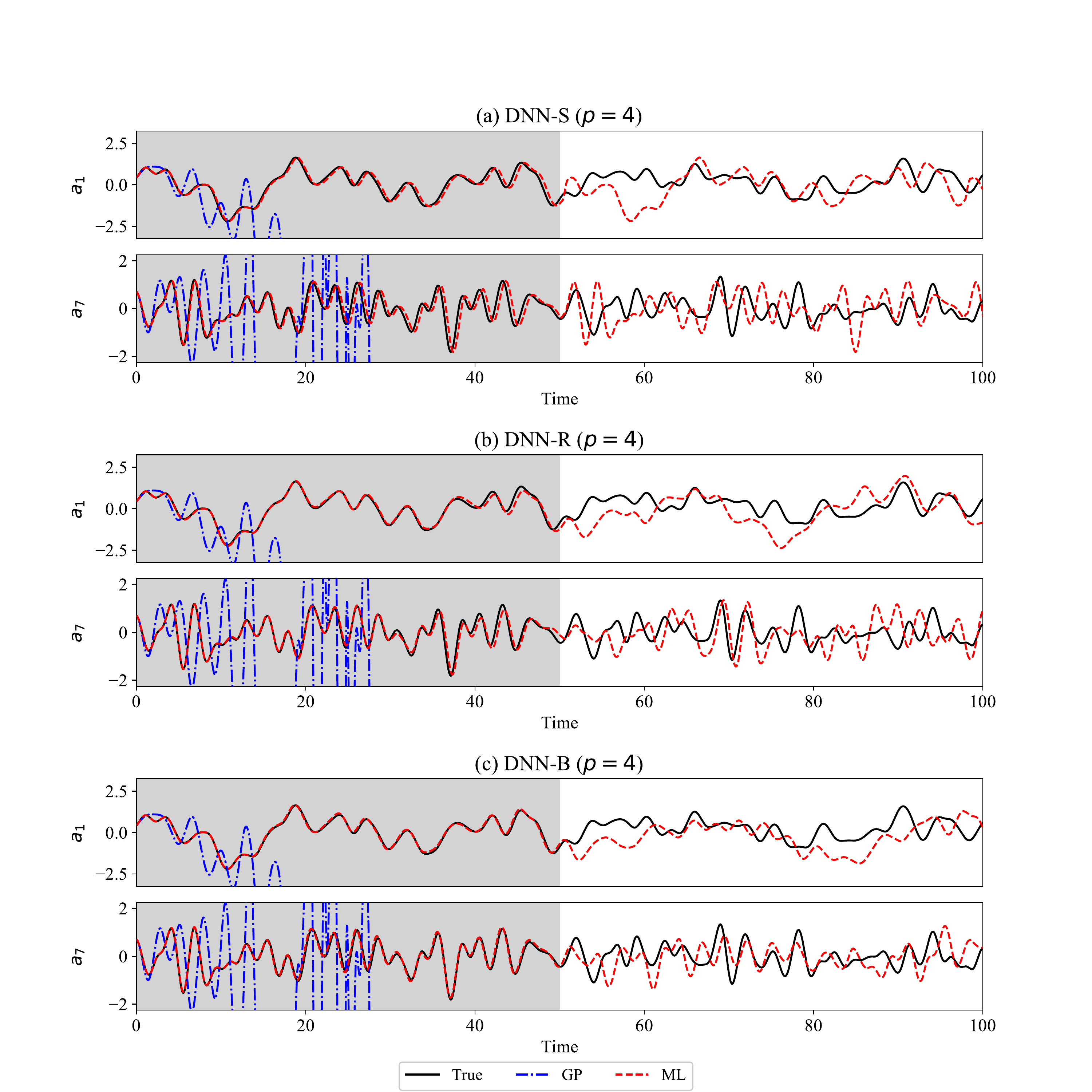}}}
\caption{Evolution of temporal coefficients for vorticity transport equation at Ra = $9.4\times10^5$ for different frameworks with $p=4$. The neural network is trained using the data highlighted in in light gray color the above figure.}
\label{fig:a94_fourleg}
\end{figure*}

The results presented in Figure \ref{fig:a34_oneleg} and Figure \ref{fig:a34_fourleg} were for Ra = $3.4\times10^5$. At lower Rayleigh number, the flow is smooth and orderly, and hence the evolution of modal coefficients was periodic. This is a simple problem with stationary time series and can be solved using simple methods like extreme learning machine\cite{huang2006extreme, san2018extreme}. The DNN will be beneficial for higher Rayleigh number case after the onset of turbulence. Figures \ref{fig:a94_oneleg} and \ref{fig:a94_fourleg} show the similar results for Ra = $9.4\times10^5$. The modal coefficients are not periodic due to the chaotic and turbulent nature of flow taking place in the cavity at such higher Rayleigh number. There is a considerable variation in the evolution of the modal coefficients as the time proceeds. The Galerkin projection is unbounded with less number of modes and gives nonphysical results for such complex flows. In order to recover the correct physics using Galerkin projection, we will have to use an increased number of modes and this will lead to increased computational cost. However, we are interested in recovering the accurate physics as much as possible with less computational cost.   

Figure \ref{fig:a94_oneleg} presents results for all DNN frameworks with $p=1$ in the input training data for Ra = $9.4\times10^5$. We can see that the DNN prediction is bounded for all DNN frameworks. \textcolor{rev1}{Although there exist necessary and sufficient conditions for global boundedness\cite{schlegel2015long} for Galerkin systems, we have not performed rigorous analysis about boundedness of the proposed frameworks. A deeper discussion of a bounded-input-bounded-output systems can be found elsewhere\cite{shi2007support}}. We observe that the DNN-R and DNN-B frameworks perform better than DNN-S framework especially in the in-sample zone (i.e., $t=0$ to $t=50$). However, we see that the DNN-R and DNN-B frameworks overpredict both modal coefficient  $a_1$ and $a_7$ for out-of-sample zone than the true projection of FOM solution on reduced order space. 
Figure \ref{fig:a94_fourleg} shows the similar results for all DNN frameworks with $p=4$ in the input training data. We notice that the prediction has improved for all DNN frameworks when we use the modal coefficient history in the input training data. All DNN frameworks are almost able to predict the true projection of FOM solution accurately in the in-sample zone with $p=4$. The problem of overprediction is also reduced for DNN-R and DNN-B frameworks by including the time history in the input data.   

Interestingly, including time history in the input training data seems to have helped neural networks to predict more accurate results for both cases (Ra = $3.4\times10^5$ and Ra = $9.4\times10^5$). One explanation for this behavior is that short-term past history of the system helps the neural network to learn the state of the system and predict future state more accurately. Our numerical experiments show that increasing the past history of the system might not help beyond some point. In some cases, including long-term history might give adverse results due to overfitting, or DNN trying to find an unrelated pattern between the input and the output.

Figures \ref{fig:a34_oneleg}-\ref{fig:a94_fourleg} show the vorticity modal coefficient only for two modes $a_1$ and $a_7$. We use total RMSE given by Equation~(\ref{eq:rmse}) to measure the quantitative performance for all DNN frameworks. The total RMSE measures the deviation between the true projection of FOM solution and the prediction by the neural network for all modes. In Table \ref{tab:rom_rmse}, we report the total RMSE for all DNN frameworks investigated in this study for vorticity and temperature modal coefficients for both Rayleigh numbers Ra = $3.4\times 10^5$ and Ra = $9.4\times 10^5$. Table \ref{tab:rom_rmse} also reports the root mean square error for ROM-G framework. It can be easily seen that all DNN frameworks perform better than ROM-G framework for both cases. At higher Rayleigh number, the ROM-G framework is unbounded and hence the error is very large. For both cases, we see an improvement in prediction in terms of RMSE for all DNN frameworks as we increase $p$ from $1$ to $4$.          

\begin{table}[htbp]
\caption{Quantitative assessment of Galerkin projection and different DNN frameworks  for vorticity and temperature modal coefficients for Ra = $3.4\times10^5$ and Ra = $9.4\times10^5$ using the total root mean square error given by Equation~(\ref{eq:rmse}).}
\begin{tabular}{p{0.2\textwidth}p{0.12\textwidth}p{0.12\textwidth}}\\
\hline\noalign{\smallskip}
\textbf{Framework} & $\text{RMSE}~({a})$ & $\text{RMSE}~({b})$\\\hline\noalign{\smallskip}
\underline{Ra = $3.4\times10^5$} & & \\
ROM-G & $9.8\times10^{-1}$ & $2.1\times10^{-2}$\\
DNN-S ($p=1$) & $9.9\times10^{-2}$ & $6.3\times10^{-3}$ \\
DNN-R ($p=1$) & $4.8\times10^{-1}$ &  $1.3\times10^{-2}$\\
DNN-B ($p=1$) & $2.9\times10^{-2}$ &  $5.3\times10^{-3}$\\
DNN-S ($p=4$) & $9.1\times10^{-2}$ &  $8.8\times10^{-4}$\\
DNN-R ($p=4$) & $1.2\times10^{-1}$ &  $2.2\times10^{-4}$\\
DNN-B ($p=4$) & $2.0\times10^{-1}$ &  $5.9\times10^{-4}$\\{\smallskip}

\underline{Ra = $9.4\times10^5$} & & \\
ROM-G & $3.1\times10^{2}$ & $1.3\times10^{0}$\\
DNN-S ($p=1$) & $8.9\times10^0$ &  $1.6\times10^{-1}$\\
DNN-R ($p=1$) & $8.3\times10^0$ &  $1.3\times10^{-1}$\\
DNN-B ($p=1$) & $7.7\times10^0$ &  $1.7\times10^{-1}$\\
DNN-S ($p=4$) & $4.9\times10^0$ &  $1.8\times10^{-1}$\\
DNN-R ($p=4$) & $5.6\times10^0$ &  $1.0\times10^{-1}$\\
DNN-B ($p=4$) & $4.6\times10^0$ &  $1.0\times10^{-1}$\\
\hline
\end{tabular}
\label{tab:rom_rmse}
\end{table}

After comparing the time evolution of vorticity, and temperature modal coefficient for all DNN frameworks, we proceed to compare the performance of proposed DNN frameworks in predicting the temperature field in the cavity at final time $t=100$. We compare our results with the true projection of FOM solution on the reduced order space. The FOM solution is obtained by DNS and is discussed briefly in Section \ref{subsec:boussinesq}. For lower Rayleigh number case, first 10 modes capture more than $99\%$ of the total energy. Hence, the FOM solution and its true projection will be close to each other. However, for higher Rayleigh number case, first 10 modes capture only 69\% of total energy and we expect to see some discrepancy between FOM solution and its true projection on reduced order space. We train the neural network using the evolution of modal coefficients for the true projection of FOM solution and hence we can recover at most true projection of FOM solution.

\begin{figure*}[htbp]
\centering
\mbox{
\subfigure{\includegraphics[width=0.9\textwidth]{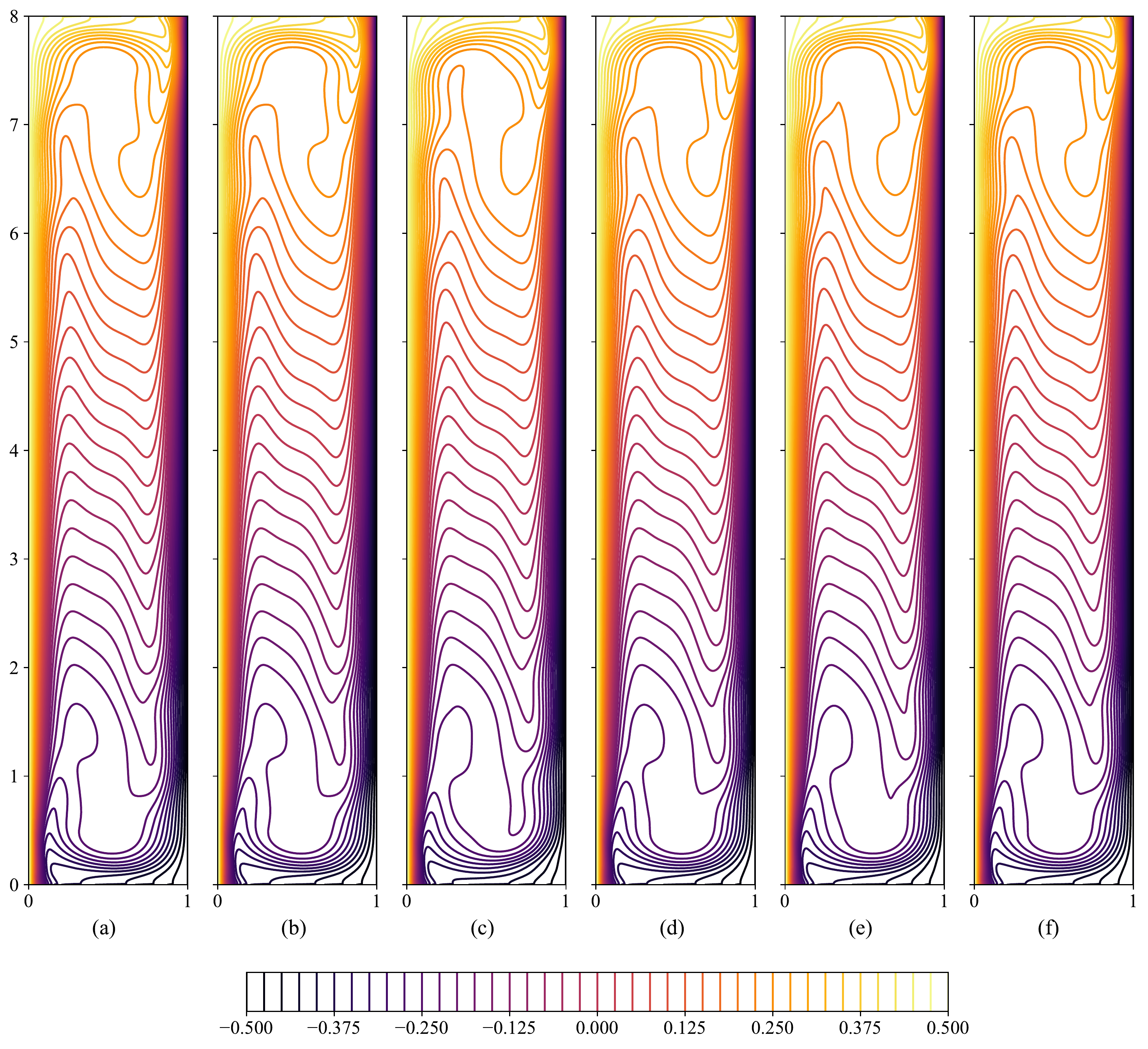}}}
\caption{Contours of instantaneous temperature at Ra = $3.4\times10^5$. The neural network is trained for different frameworks with $p=1$. (a) FOM, (b) True projection, (c) ROM-G, (d) DNN-S framework, (e) DNN-R framework, (f) DNN-B framework.}
\label{fig:a34_cont_oneleg}
\end{figure*}

\begin{figure*}[htbp]
\centering
\mbox{
\subfigure{\includegraphics[width=0.9\textwidth]{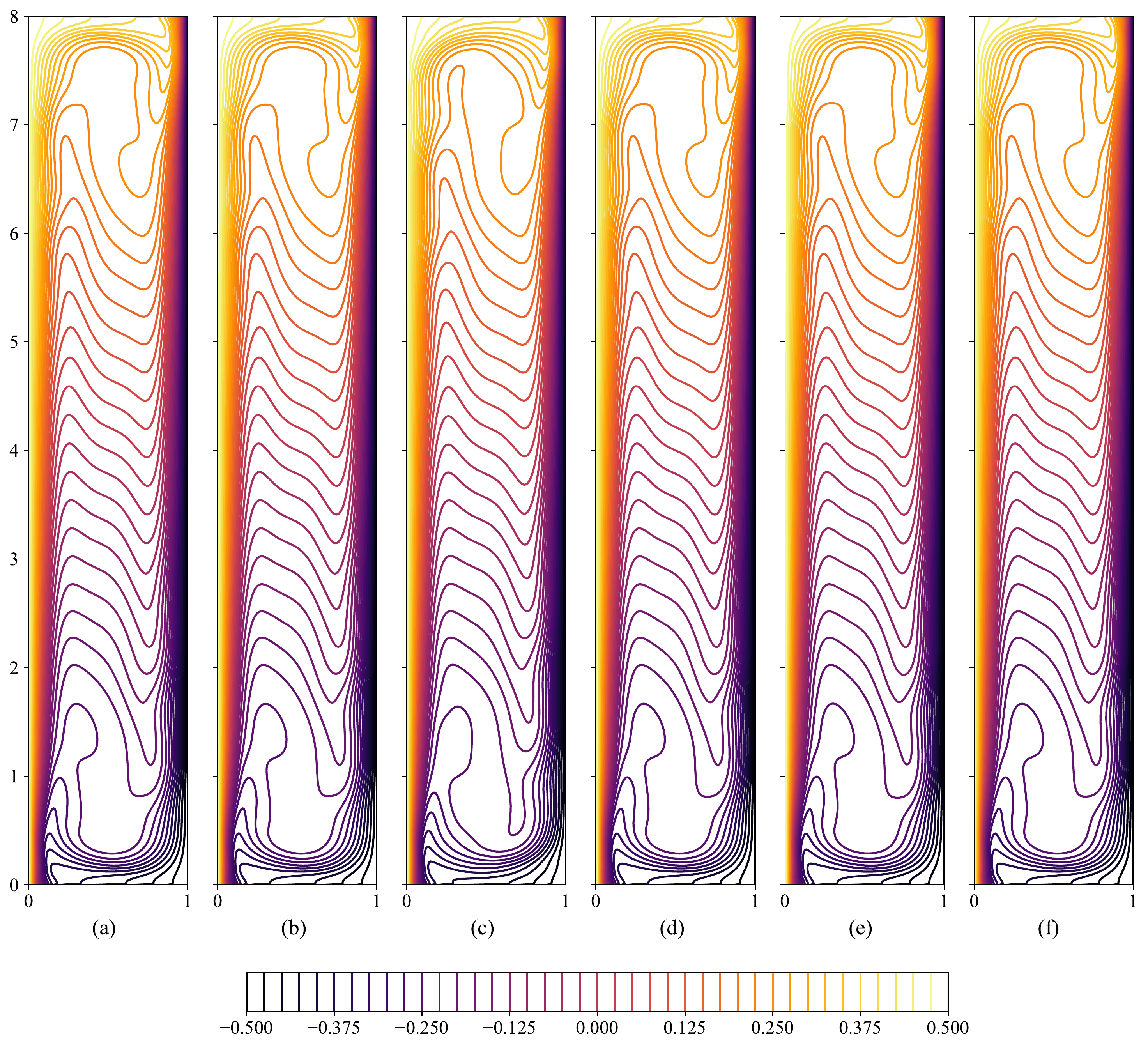}}}
\caption{Contours of instantaneous temperature at Ra = $3.4\times10^5$. The neural network is trained for different frameworks with $p=4$. (a) FOM, (b) True projection, (c) ROM-G, (d) DNN-S framework, (e) DNN-R framework, (f) DNN-B framework.}
\label{fig:a34_cont_fourleg}
\end{figure*}

We compute the instantaneous temperature field using Equation~(\ref{eq:deco}). The coefficient $b_k$ is considered at the final time step $t=100$. Figure \ref{fig:a34_cont_oneleg} displays the temperature field at the final time predicted by FOM simulation, true projection of FOM on reduced order space, ROM-G framework, and all DNN frameworks. We observe that the FOM solution and its true projection are almost identical. We see that there is some deviation in the temperature field predicted using ROM-G framework and the true projection. The temperature field predicted using DNN-S and DNN-B frameworks are very close to the true projection solution. We see some variation in the solution predicted by DNN-R framework, which can be attributed to the phase shift in modal coefficient predicted by DNN-R framework with $p=1$ (as can be seen in Figure \ref{fig:a34_oneleg}). Figure \ref{fig:a34_cont_fourleg} shows the similar results for lower Rayleigh number case with $p=4$. We notice an improvement in prediction by the DNN-R framework and the predicted temperature field is close to the true projection solution.  

\begin{figure*}[htbp]
\centering
\mbox{
\subfigure{\includegraphics[width=0.9\textwidth]{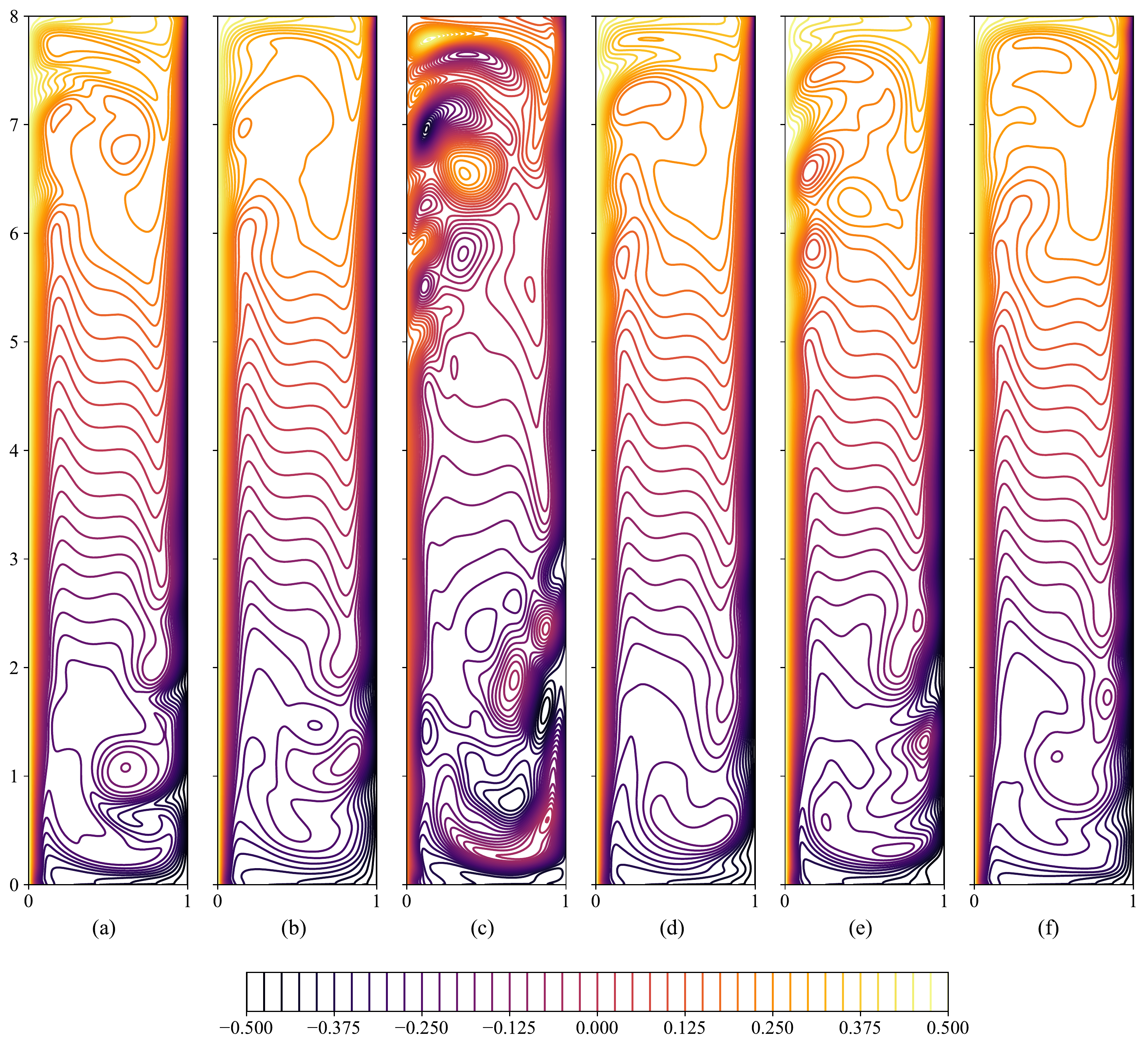}}}
\caption{Contours of instantaneous temperature at Ra = $9.4\times10^5$. The neural network is trained for different frameworks with $p=1$. (a) FOM, (b) True projection, (c) ROM-G, (d) DNN-S framework, (e) DNN-R framework, (f) DNN-B framework.}
\label{fig:a94_cont_oneleg}
\end{figure*}

\begin{figure*}[htbp]
\centering
\mbox{
\subfigure{\includegraphics[width=0.9\textwidth]{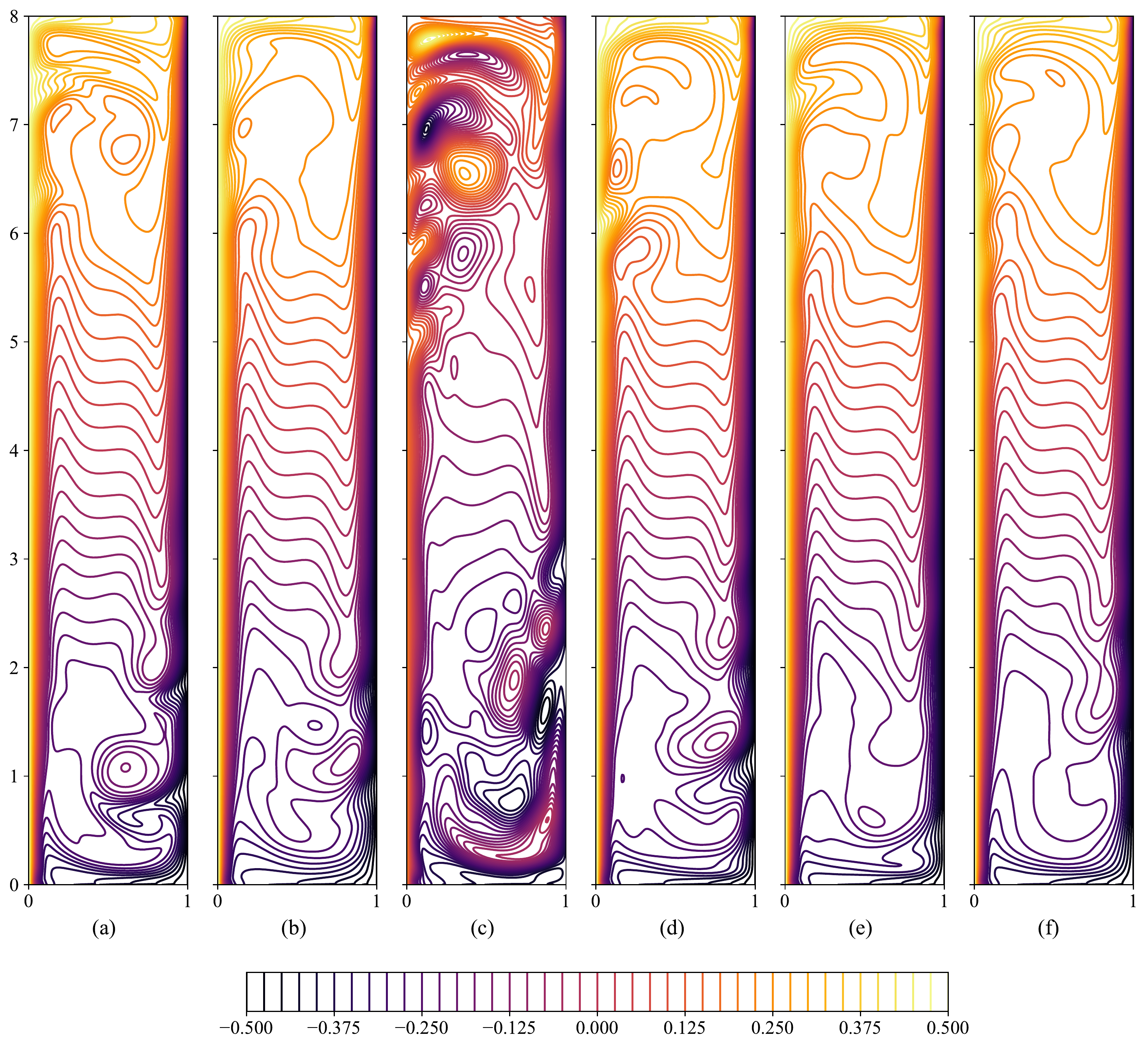}}}
\caption{Contours of instantaneous temperature at Ra = $9.4\times10^5$. The neural network is trained for different frameworks with $p=4$. (a) FOM, (b) True projection, (c) ROM-G, (d) DNN-S framework, (e) DNN-R framework, (f) DNN-B framework.}
\label{fig:a94_cont_fourleg}
\end{figure*}

Figure \ref{fig:a94_cont_oneleg} illustrates the similar results for temperature field at final time $t=100$ for higher Rayleigh number case. The neural network is trained using $p=1$ in the input training data. We see some of the discrepancies between the FOM solution and its projection on reduced order space. The discrepancy is due to less amount of energy in the first 10 POD modes, and hence, some of the flow features get neglected. The deviation is mainly observed at bottom and top of the heated cavity due to the vortical structures formed in these regions at higher Rayleigh number. From Figure \ref{fig:a94_cont_oneleg} we see that the solution predicted by ROM-G framework is very different from the true projection for higher Rayleigh number case. The solution predicted by all DNN frameworks is not identical to the true projection solution. The difference is primarily seen at the bottom and top region. However, we see overall good qualitative agreement between the temperature field predicted by DNN and true projection solution. Figure \ref{fig:a94_cont_fourleg} shows similar results for higher Rayleigh number case when the neural network is trained using $p=4$ in the input training data. We observe a slight improvement in the temperature field prediction. This is consistent with an improvement in the prediction of modal coefficient (refer to Figure \ref{fig:a94_fourleg}) with an increase in the solution history in the input data for the neural network. The prediction of modal coefficients is very close to the modal coefficients for the true projection of FOM solution in the in-sample zone ($t=0$ to $t=50$). We do not get the similar accuracy for the out-of-sample zone ($t=50$ to $t=100$). Therefore, we are not able to recover the true projection solution exactly. Despite its limitations, it is clear that our data-driven non-intrusive framework is robust and accurate compared to the intrusive ROM-G framework, and can be used for challenging flow problems where the flow is turbulent and not uniform.         

Next, we calculate the time-averaged Nusselt number along the left wall ($x=0$). This quantity can be of interest in many engineering applications. The instantaneous Nusselt number is calculated using Equation~(\ref{eq:nusselt}). The temperature in Equation~(\ref{eq:nusselt}) can be obtained using Equation~(\ref{eq:deco}). The gradient of the temperature is computed using a right-sided second-order finite difference scheme. The integration of the temperature gradient is computed using Simpson's integration rule. After calculating the instantaneous temperature at every time step, we take its average to get the time-averaged Nusselt number. The detail formulas for calculating the instantaneous Nusselt number are provided in the appendix.    

\begin{figure*}[htbp]
\centering
\mbox{
\subfigure{\includegraphics[width=0.9\textwidth]{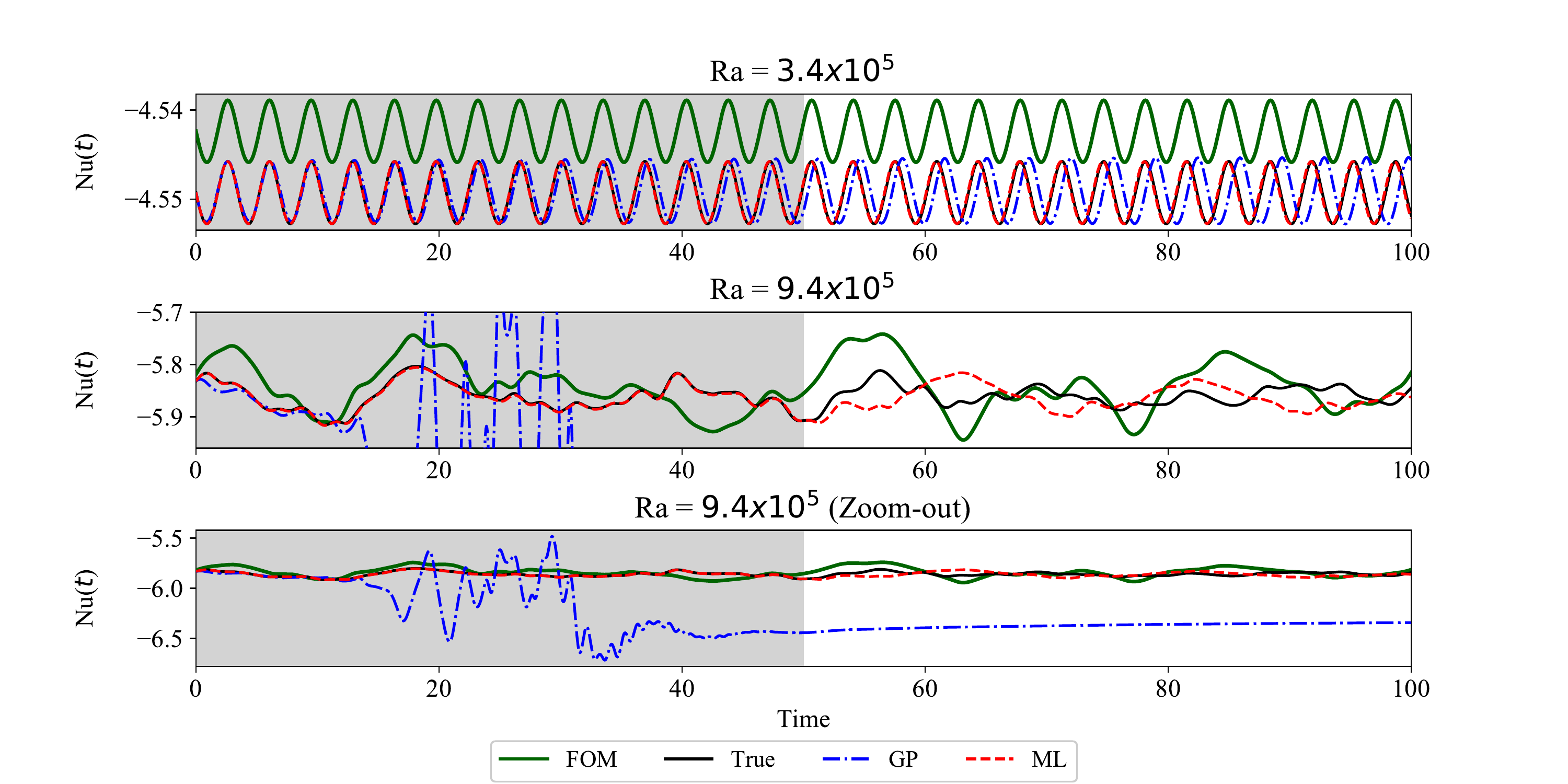}}}
\caption{Evolution of instantaneous Nusselt number for two different Rayleigh numbers for FOM, True projection, ROM-G, and DNN-R framework with $p=4$. The neural network is trained using the data highlighted in light gray color in the above figure. The bottom figure shows the zoom-out plot for the middle figure to show the large range of variation in Nusselt number prediction. }
\label{fig:nu_dnn_r}
\end{figure*}

In Table \ref{tab:nusselt}, we list the statistics of time-averaged Nusselt number for FOM solution, true projection of FOM solution, ROM-G framework, and all DNN frameworks investigated in this study. We can notice all DNN frameworks and intrusive ROM-G framework gives an accurate prediction of the time-averaged Nusselt number. The standard deviation of Nusselt number is also predicted correctly by all DNN frameworks and ROM-G framework for lower Rayleigh number case. We do not recover similar results for higher Rayleigh number case. The time-averaged Nusselt number for true projection solution is slightly different from the FOM solution. The ROM-G framework overpredicts the time-averaged Nusselt number and the difference between the true projection solution and ROM-G prediction is significant. On the other hand, all DNN frameworks predicted the time-averaged Nusselt number close to the true projection solution. The standard deviation of the Nusselt number is also predicted with sufficient accuracy for all DNN frameworks. 

To illustrate the temporal variation of Nusselt number, we show the evolution of Nusselt number for FOM, true projection, ROM-G, and DNN-R framework for both Rayleigh numbers in Figure~\ref{fig:nu_dnn_r}. It can be clearly seen that the ROM-G framework fails to predict the temporal behavior of the Nusselt number correctly especially at Ra = $9.4\times10^5$. At lower Rayleigh number, Ra = $3.4\times10^5$, there is a phase shift in the Nusselt number prediction by the ROM-G framework. At this Rayleigh number, with limit cycle oscillations, the Nusselt number is slightly underpredicted (up to second digit accurate) by the true projection of FOM solution on reduced order bases. All DNN frameworks predict the Nusselt number accurately close to the true projection results. We would like to again emphasize that the neural network is trained using the true projection of FOM solution and hence we can at most recover the true projection results and not the FOM results. We also note that much simpler methods, like autocorrelation analysis (instead of the heavy DNN), can be used in predicting this stationary time series problem \cite{shumway2017time}. For higher Rayleigh number, the Nusselt number prediction for the ROM-G framework becomes unbounded in the beginning and then it calculates the overpredicted value of instantaneous Nusselt number similar to modal coefficients for vorticity and temperature field. The DNN-R framework correctly predicts the instantaneous Nusselt number close to the true projection of the FOM solution in the in-sample zone and sufficiently accurate results for the out-of-sample zone. To avoid redundancy, we do not present results for other DNN-frameworks since we get a similar prediction.  

\begin{table}[htbp]
\caption{Statistics of Nusselt number computed on the left wall ($x=0$) for Ra = $3.4\times10^5$ and Ra = $9.4\times10^5$. The mean Nusselt number and its standard deviation is computed from instantaneous Nusselt number from time period $t=0$ to $t=100$.}
\begin{tabular}{p{0.2\textwidth}p{0.12\textwidth}p{0.12\textwidth}}\\
\hline\noalign{\smallskip}
\textbf{Framework} & ${\mu}$ & ${\sigma}$\\\hline\noalign{\smallskip}
\underline{Ra = $3.4\times10^5$} & & \\
FOM & -4.5425 & $2.46\times 10^{-3}$\\
True & -4.5494 & $2.48\times 10^{-3}$\\
ROM-G & -4.5492 & $2.55\times 10^{-3}$\\
DNN-S ($p=1$) & -4.5494 &  $2.48\times 10^{-3}$\\
DNN-R ($p=1$) & -4.5494 &  $2.50\times 10^{-3}$\\
DNN-B ($p=1$) & -4.5494&  $2.49\times 10^{-3}$\\
DNN-S ($p=4$) & -4.5494 &  $2.48\times 10^{-3}$\\
DNN-R ($p=4$) & -4.5494 &  $2.48\times 10^{-3}$\\
DNN-B ($p=4$) & -4.5494 &  $2.48\times 10^{-3}$\\{\smallskip}

\underline{Ra = $9.4\times10^5$} & & \\
FOM & -5.8411 & $4.81\times10^{-2}$\\
True & -5.8603 & $2.29\times10^{-2}$ \\
ROM-G & -6.2530 & $2.43\times10^{-1}$\\
DNN-S ($p=1$) & -5.8572 & $3.45\times10^{-2}$ \\
DNN-R ($p=1$) &  -5.8643 & $2.14\times10^{-2}$\\
DNN-B ($p=1$) & -5.8639 &  $2.19\times10^{-2}$\\
DNN-S ($p=4$) & -5.8634 & $2.87\times10^{-2}$ \\
DNN-R ($p=4$) & -5.8637 & $2.50\times10^{-2}$ \\
DNN-B ($p=4$) & -5.8642 & $2.29\times10^{-2}$ \\
\hline
\end{tabular}
\label{tab:nusselt}
\end{table}

To summarize, we demonstrated the capability of our DNN frameworks within the non-intrusive ROM setup for the differentially heated cavity problem at two Rayleigh numbers. We do a systematic analysis of our DNN frameworks in terms of prediction of the time evolution of modal coefficients, instantaneous temperature field prediction at the final time, and prediction of time-averaged quantities. Our non-intrusive ROM framework gives sufficiently accurate results for simple as well as complex flows.   

\section{Concluding Remarks} 
\label{sec:conclusion}
In this work, we put forward a non-intrusive reduced order modeling framework which uses a deep neural network to predict the dynamics of ROM for complex flow problems. Deep neural networks are capable of approximating a complex nonlinear relationship between the input and output and we achieve this via supervised learning task.  The key difference between proposed DNN frameworks is the output variable that is learned by the neural network. The non-intrusive ROM is devised using an encoder-decoder approach and DNN is used for predicting the modal coefficients in an iterative fashion. We use our DNN frameworks with multiple temporal legs (short term history of the state of the system) in the input data. This enables the neural network to take the memory effect into account for predicting the future state of the system. We leverage the classical numerical schemes (backward-difference) in our proposed DNN-B framework, and this framework can also be implemented with other numerical schemes such as Adams-Bashforth, and Adams-Moulton families. First, we use all DNN frameworks for two benchmark problems: Kraichnan-Orszag system and Lorenz system. All DNN frameworks are able to correctly predict each state of the Kraichnan-Orszag system. Even though all DNN frameworks fail to predict correct trajectories for each state of the Lorenz system for a longer duration of time, it predicts the correct dynamics of Lorenz system in terms of the Lorenz attractor.  

After evaluating all DNN frameworks for predicting the dynamics of nonlinear dynamical systems, we proceed to reduce order modeling of the differentially heated cavity problem. We extract 1000 snapshots from DNS simulation after the steady state has been reached. The POD bases are constructed using these 1000 snapshots. Based on existing literature and our findings from DNS simulation, we see the change from periodic flow to turbulent flow with an increase in Rayleigh number. For this reason, we test our proposed non-intrusive ROM framework for two cases: lower Rayleigh number case (Ra = $3.4\times 10^5$) and higher Rayleigh number case (Ra = $9.4\times 10^5$). We use 10 POD modes for our analysis. Due to the turbulent nature of flow at higher Rayleigh number, the first 10 modes capture only 69\% of the energy. We sacrifice the advantage of ROM if we increase the number of modes ($R=40$ for 95\% of the energy) and hence we attempt to get results close to the true projection of FOM solution on reduced order space with the proposed non-intrusive ROM framework. We assess the performance of non-intrusive ROM framework using different parameters such as prediction capability of modal coefficients, prediction of instantaneous temperature at the final time, and prediction of engineering quantities of interest such as time-averaged Nusselt number. The non-intrusive ROM frameworks perform exceptionally well for all these parameters for lower Rayleigh number case. We see some deviation with the prediction of modal coefficients for higher Rayleigh number case (especially for the out-of-sample zone). Despite this deviation, the proposed framework is able to predict the instantaneous temperature field and time-averaged Nusselt number with sufficient accuracy.        

We also compared our results with the results for Galerkin projection (ROM-G framework). The ROM-G framework gives a good prediction for low Rayleigh number case. However, the ROM-G framework is unbounded for higher Rayleigh number case, and produces the wrong prediction. Our analysis for differentially heated cavity problem at two Rayleigh numbers indicates that the non-intrusive ROM setup equiped with the DNN frameworks yields satisfactorily accurate results, and has a potential for reduced order modeling of complex flow problems. \textcolor{rev1}{In this paper, we used snapshot POD to represent the high-dimensional data onto low-dimensional space. At this point, we can ask the following question: is the POD preconditioning for the model identification step necessary? Given the analogy between POD and shallow autoencoder, it is possible to use deep neural networks to provide a more compact representation of high-dimensional data\cite{brunton2019machine}. We might arguably assume that a DNN could also learn the potentially much lower dimensional manifold from snaphsot data\cite{loiseau2018sparse}. Indeed, this is an exciting future direction and the work is in progress on this topic. }

\begin{acknowledgements}
This material is based upon work supported by the U.S. Department of Energy, Office of Science, Office of Advanced Scientific Computing Research under Award Number DE-SC0019290. O.S. gratefully acknowledges their support. Direct numerical simulations for this project were performed using resources of the Oklahoma State University High Performance Computing Center. We also gratefully acknowledge the support of NVIDIA Corporation with the donation of the GeForce Titan Xp GPU for our research. Disclaimer: This report was prepared as an account of work sponsored by an agency of the United States Government. Neither the United States Government nor any agency thereof, nor any of their employees, makes any warranty, express or implied, or assumes any legal liability or responsibility for the accuracy, completeness, or usefulness of any information, apparatus, product, or process disclosed, or represents that its use would not infringe privately owned rights. Reference herein to any specific commercial product, process, or service by trade name, trademark, manufacturer, or otherwise does not necessarily constitute or imply its endorsement, recommendation, or favoring by the United States Government or any agency thereof. The views and opinions of authors expressed herein do not necessarily state or reflect those of the United States Government or any agency thereof.
\end{acknowledgements}

\setcounter{equation}{0}
\renewcommand{\theequation}{A\arabic{equation}}

\section*{Appendix: Nusselt number calculation}
\label{app:nusselt}
The instantaneous Nusselt number of the flow  along the left wall ($x=0$) is given by
\begin{equation}
    \text{Nu($t$)} = \frac{1}{H} \int_{0}^{H} \frac{\partial \theta}{\partial x} \Big|_{x=0}dy,
    \label{eq:app_nusselt}
\end{equation}
where $H$ is the height of the cavity.

For the FOM simulation, we have the temperature data available for each discrete point at each time step. The gradient of the temperature is computed using the right-sided finite difference scheme along the left wall ($x=0$) as given by the below equation 
\begin{equation}\label{eq:app_fdm}
    \frac{\partial\theta}{\partial x} \Big|_{x=0} = \frac{-3\theta_{0,j} + 4\theta_{1,j}-\theta_{2,j}}{2h},    
\end{equation}
where $j$ represents the discrete spatial location in $y$-direction, and $h$ is the grid spacing in $x$-direction. The numerical integration is performed using the fourth-order accurate Simpson's rule. The mean and the standard deviation of the Nusselt number are then computed from the series of instantaneous Nusselt number evaluated at each time step between $t=0$ to $t=100$ using below formulas
\begin{align}
    \mu &= \frac{1}{N}\sum_{n=1}^{N} \text{Nu($t_n$)}, \label{eq:mean}\\
    \sigma &= \sqrt{\frac{1}{N-1}\sum_{n=1}^{N} (\text{Nu($t_n$)}-\mu)^2}, \label{eq:std}
\end{align}
where $N$ is the total number of time steps between $t=0$ to $t=100$. 

In the ROM framework, the full order solution can be recovered using the modal coefficient using Equation~(\ref{eq:deco}) and is also given below
\begin{equation}\label{eq:app_temp}
    \theta(\textbf{x}, t) = \bar{\theta}(\textbf{x}) + \sum_{k=1}^{R} b_{k}(t)\phi_k^{\theta}(\textbf{x}),
\end{equation}
where $R$ is the number of modes retained for POD ($R=10$ in this study). 

We can compute the temperature field at each time step using Equation~(\ref{eq:app_temp}) and then follow the same procedure similar to FOM solution. Another faster method to determine the Nusselt number is to calculate the gradient of the mean temperature and each basis function and store it in the memory. The stored gradient can be used to find the instantaneous temperature gradient using equation given below
\begin{equation}\label{eq:app_gradtemp}
    \frac{\partial\theta}{\partial x}\Big|_{x=0} = \frac{\partial \bar{\theta}}{\partial x}\Big|_{x=0} + \sum_{k=1}^{R} b_{k}(t) \frac{\partial \phi_k^{\theta}}{\partial x}\Big|_{x=0}.
\end{equation}
If we substitute Equation~(\ref{eq:app_gradtemp}) in Equation~(\ref{eq:app_nusselt}) we get
\begin{equation}
    \text{Nu($t$)}=\kappa+\sum_{k=1}^{R} b_{k}(t)\gamma_{k},
\end{equation}
where the predetermined coefficients are 
\begin{align}
    \kappa=\frac{1}{H} \int_{0}^{H}\frac{\partial \bar{\theta}}{\partial x}\Big|_{x=0}dy, \\
    \gamma_k=\frac{1}{H} \int_{0}^{H} \frac{\partial \phi_k^{\theta}}{\partial x}\Big|_{x=0}dy.
\end{align}
The gradient for the mean temperature and basis function is evaluated using Equation~(\ref{eq:app_fdm}). The temporal value of temperature modal coefficient is determined using Galerkin projection for ROM-G framework and using DNN frameworks for non-intrusive ROM setup. The numerical integration of instantaneous temperature gradient along the left wall is computed using the composite Simpson's rule with SciPy function  \texttt{integrate.simps} for all ROMs. The mean and standard deviation of the Nusselt number is then computed using equations~(\ref{eq:mean}) and (\ref{eq:std}), respectively.  

\bibliography{reference}

\end{document}